\documentclass[a4paper]{llncs} %[runningheads,a4paper]
\usepackage{amssymb}
\usepackage{graphicx}
\usepackage{url}
\usepackage{multirow}
\usepackage{verbatim}
\usepackage[english]{babel}
\usepackage{rotating}
\usepackage{authblk}
\usepackage{alltt}
\usepackage{moreverb}
\usepackage{theorem}
\usepackage{amsmath}
\usepackage{multirow}
\usepackage[table]{xcolor}
\usepackage{array,ragged2e}
\usepackage{longtable}
\usepackage{pdflscape}

\usepackage{lipsum}
\usepackage{booktabs}
\usepackage{dblfloatfix}
\usepackage{multirow}
\usepackage{verbatim}
\usepackage{alltt}
\usepackage{afterpage}

\usepackage{caption}

\usepackage{pgf-pie}

\usepackage{tikz}

\usetikzlibrary{positioning,shadows}

\newif\ifpienumberinlegend
\pgfkeys{/number in legend/.code=
    \expandafter\let\expandafter\ifpienumberinlegend
    \csname if#1\endcsname
    \ifpienumberinlegend

    \def\beforenumber##1\afternumber{}%
    \fi,
    /number in legend/.default=true
}
\begin{document}

\mainmatter  % start of an individual contribution

% first the title is needed
\title{Ontologies for Privacy Requirements Engineering: A Systematic Literature Review}

% a short form should be given in case it is too long for the running head
\titlerunning{Ontologies for Privacy Requirements Engineering: A Systematic Literature Review}

% the name(s) of the author(s) follow(s) next
%
% NB: Chinese authors should write their first names(s) in front of
% their surnames. This ensures that the names appear correctly in
% the running heads and the author index.
%

\author{Mohamad Gharib\inst{1} \and    Paolo Giorgini\inst{2}  \and    John Mylopoulos\inst{2}}

\authorrunning{Gharib et al.}

% (feature abused for this document to repeat the title also on left hand pages)

% the affiliations are given next; don't give your e-mail address
% unless you accept that it will be published

\institute{University of Florence - DiMaI, Viale Morgagni 65, Florence, Italy \\   \email{mohamad.gharib@unifi.it}
\and     University of Trento - DISI, 38123, Povo, Trento, Italy\\    	\email{\{paolo.giorgini,john.mylopoulos\}@unitn.it}}

%
% NB: a more complex sample for affiliations and the mapping to the
% corresponding authors can be found in the file "llncs.dem"
% (search for the string "\mainmatter" where a contribution starts).
% "llncs.dem" accompanies the document class "llncs.cls".
%

\toctitle{Lecture Notes in Computer Science}
\tocauthor{Authors' Instructions}
\maketitle

\begin{abstract}

Privacy has been frequently identified as a main concern for system developers while dealing with/managing personal information. Despite this, most existing work on privacy requirements deals with them as a special case of security requirements. Therefore, key aspects of privacy are, usually, overlooked. In this context, wrong design decisions might be made due to  insufficient understanding of privacy concerns. In this paper, we address this problem with a systematic literature review whose main purpose is to identify the main concepts/relations for capturing privacy requirements.  In addition, the identified concepts/relations are further analyzed to propose a novel privacy ontology to be used by software engineers when dealing with privacy requirements.

\section*{Keywords} Privacy Ontology,  Privacy Requirements, Privacy by Design (PbD), Requirements Engineering

\end{abstract}

\section{Introduction}

Increasing numbers of today's systems deal with personal information (e.g., information about citizens, customers, etc.), where such information is protected by privacy laws \cite{gharibre2016}.  Therefore, privacy has become a main concern for system designers. In other words, dealing with privacy related concerns is a must these days because privacy breaches may result in huge costs as well as a long-term consequences \cite{acquisti2006there,gellman2002privacy,hong2004privacy,camp2002designing,campbell2003economic}. Privacy breaches might be due lack of appropriate security policies, bad security practices,  attacks, data thefts, etc.  \cite{acquisti2006there,labda2014modeling}. However, most of these breaches can be avoided if privacy requirements of the system-to-be were captured properly during system design (e.g., Privacy by Design (PbD)) \cite{cavoukian2009privacy,cavoukian2011privacy,labda2014modeling}, where privacy requirements aim to capture the types and levels of protection necessary to meet the privacy needs of the users.

Nevertheless, just  few works focused on considering privacy during the system design \cite{Gurses2011}. More specifically, most existing work on privacy requirements often deal with them either as non-functional requirements (NFRs) with no specific criteria on how such requirements can be met \cite{anton2002analyzing,yu2002designing,mouratidis2007secure}, or as a part of security \cite{zannone2006requirements,kalloniatis2008addressing}, i.e.,  focusing mainly on confidentiality and overlooking important privacy aspects such as anonymity, pseudonymity, unlinkability, unobservability, etc.

On the other hand,  privacy is an elusive and vague concept \cite{solove2002conceptualizing,solove2006taxonomy,kalloniatis2008addressing}. Although several efforts have been made to clarify the privacy concept by linking it to more refined concepts such as secrecy, person-hood, control of personal information, etc., there is no consensus on the definition of these concepts or which of them should be used to analyze privacy \cite{solove2006taxonomy}. This has resulted in a lot of confusion among designers and stakeholders, which has led in turn to wrong design decisions. In this context, a well-defined privacy ontology that captures privacy related concepts along with their interrelations would constitute a great step forward in designing privacy-aware systems.  

Ontologies have been proven to be a key success factor for eliciting high-quality requirements, and it can facilitate and improve the job of requirements engineers \cite{souag2012towards,kaiya2006using,dzung2009ontology}, since it can reduce the conceptual vagueness and terminological confusion by providing a shared understanding of the related concepts between  designers and stakeholders \cite{uschold1996ontologies}.  

In addition, the ontology should capture  privacy requirements in their social and organizational context. Since most complex systems these days (e.g., healthcare systems, smart cities, etc.) are socio-technical systems \cite{emery1960socio}, which consist not only of technical components but also of humans along with their interrelations, where different kinds of vulnerabilities might manifest themselves \cite{liu2003security,gharibre2016}. Focusing on the technical aspects and leaving the social and organizational aspects outside the system's boundary leaves the system open to different kinds of vulnerabilities that might manifest themselves in the social interactions and/or the organizational structure of the system \cite{liu2003security}. The Flash Crash \cite{sommerville2012large} and the Allied Irish Bank scandal \cite{massacci2008detecting} are good examples, where  problems were not caused by mere technical failures, but it were also due to several socio-technical related vulnerabilities of the system.

This paper applies systematic review techniques to survey available  literature to identify the most mature studies that propose privacy ontologies/concepts.  In addition, we further analyze the selected privacy related concepts/relations to identify the main ones in order to propose a novel ontology that can be used to capture privacy requirements.   This paper is therefore intended to be a starting point to address the problem of identifying a core privacy ontology.

The rest of the paper is organized as follows; Section (\textsection 2) describes the review process and the protocol underlining  this systematic review. We present and discuss the review results and findings in Section (\textsection 3). In Section (\textsection 4) we propose a novel ontology for privacy requirements engineering.  We discuss threats to validity in Section (\textsection 5). Related work is presented in Section (\textsection 6), and we conclude and discuss future work in Section (\textsection 7).

\section{Review Process}

\begin{figure*}[!t]
\centering
\includegraphics[width= 0.99 \linewidth]{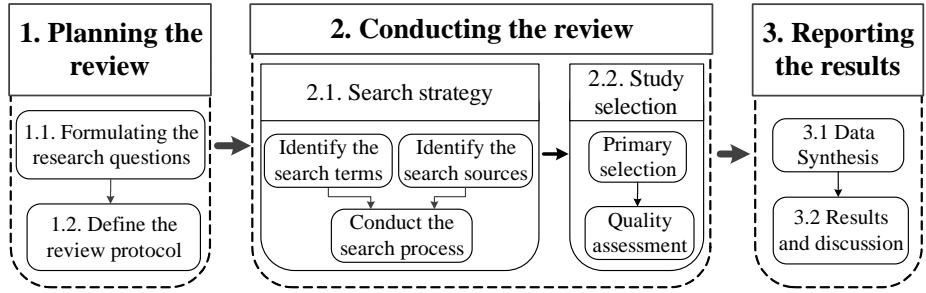}
\caption{The systematic review process}
%\vskip -6pt
\label{fig:plan}
\end{figure*}

A systematic review can be defined as a systematic process for defining research questions, searching the literature for the best available resources to answer such questions, and collecting available data from the resources for answering the research questions. Following \cite{kitchenham2004procedures,keele2007guidelines}, the review process (depicted in Figure \ref{fig:plan}) consists of three main phases: 

\begin{enumerate}

\item  Planning the review, in which we formulate the research questions and we define the review protocol.

\item Conducting the review, in which we conduct the search process after identifying the search terms and the literature sources, and then we perform the study selection activity.

\item  Reporting the results of the review, in which we collect detailed information from the selected studies in order to answer the research questions, and then we use the obtained data to answer the research questions, which we discuss in the following section.

\end{enumerate}

\subsection{Planning the review}

This phase is very important for the success of the review, for it is here that we define the research objectives and the way in which the review will be carried out. This includes two main activities: (1) formulating the research questions that the systematic review will answer; and (2) defining the review protocol that specifies the main  procedures to be taken during the review.

\subsubsection{Research questions}

Formulating the review questions is a very critical activity since these questions are used to derive the entire systematic review methodology \cite{kitchenham2004procedures}.  Therefore, we formulate the following four Research Questions (RQ) to identify the main privacy concepts that have been presented in the literature:

\begin{description}
\item[RQ1] What are the privacy concepts/relations that have been used to capture privacy concerns?
\item[RQ2] What are the main concepts/relations that have been used for capturing privacy requirements? 
\item[RQ3] Do existing privacy studies cover the main privacy concepts/relations?
\item[RQ4] What are the limitations of existing privacy studies?
\end{description}

\subsubsection{Define the review protocol}

The review protocol specifies the methods to be followed while conducting the systematic review. Based on \cite{kitchenham2004procedures,keele2007guidelines}, a review protocol should specify the following: the strategy that will be used to search for primary studies selection; study selection criteria; study quality assessment criteria; data extraction and dissemination strategies. In the rest of this section, we  discuss how we specify and perform each of these activities.

\subsection{Conducting the review}
This phase is composed of two main activities: 1- search strategy; and 2- study selection, where each of them is composed of several sub-activities. In what follows, we discuss them.

\subsubsection{Search strategy}

The search strategy aims to find as many studies relating to the research questions as possible using an objective and repeatable search strategy \cite{kitchenham2004procedures}. The search  activity consists of three main sub-activities: 1- identify the search terms, 2- identify the literature resources, and 3- conduct the search process.

\textbf{Identify the search terms.} Following \cite{kitchenham2004procedures,keele2007guidelines}, we derived the main search terms from the research questions. In particular,  we used the Boolean AND to link the major terms, and we use the Boolean OR to incorporate alternative synonyms of such terms. The resulting search terms are:  (Privacy AND (ontology OR ontologies OR taxonomy OR taxonomies ) OR (Privacy requirements).

\textbf{Identify  the literature resources.} Six electronic database resources were used to primarily extract data for this research. These include: IEEE Xplore, ACM Digital Library,  Springer, ACM library, Google Scholar, and Citseerx. 

\begin{figure}[!t]
\centering
\includegraphics[width= 0.75 \linewidth]{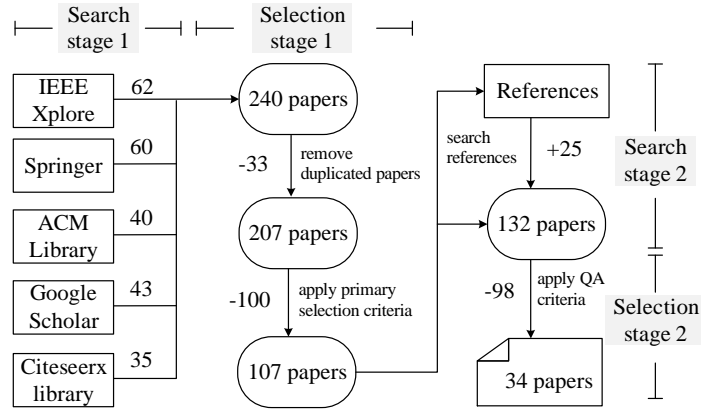}
\caption{Paper search and selection process}
%\vskip -6pt
\label{fig:process}
\end{figure}

\textbf{Conduct the search process.} The search process (shown in Figure \ref{fig:process}) consists of two main stages: 

\begin{description}
\item[Search stage 1.] We have used the search terms to search the six electronic database sources, and only papers with relevant titles have been selected; 
\item[Search stage 2.] The reference lists of all primary selected papers were carefully checked, and several relevant papers (25 papers) were identified and added to the list of the primary selected papers. 
\end{description}

\textbf{Study selection.} The selection process (shown in Figure \ref{fig:process}) consists of two main stages. 

\begin{description}
\item[Selection stage 1 (primary selection).] Searching the electronic database source returned 240 relevant papers, among which we have identified and removed 33 duplicated papers.  Next, we have applied the primary selection criteria on the remaining 207 papers. In particular, we have read the abstract, introduction, and then we skimmed through the rest of paper. We removed all the papers that are not published in the English language, and we excluded all papers that are not related to any of our research questions. Moreover, when we were able to identify multiple version of the same paper, only the most complete one was included. Finally, we excluded any paper that has been published before 1996, since we were not able to find any concrete work related to our research before 1996. The primary selection inclusion and exclusion criteria are shown in Table \ref{table:inexcriteria}. The outcome of this selection stage was 107 papers, i.e., we have excluded 100 papers.

\item[Selection stage 2 (Quality Assessment (QA)).]  At this stage, the QA criteria has been applied to the papers that have resulted from the first selection stage (107 papers) along with the papers that have resulted  from the second search stage (25 papers), for a total of 132 papers. In order to identify the most relevant studies that can be used to answer our research questions, we formulated five QA questions (shown in Table \ref{table:qualityassessment}) to evaluate the relevance, completeness, and quality of the studies, where each question has only two answers: Yes = 1 or No= 0. The quality score for each study is computed by summing the scores of its QA questions, and the paper is selected only if it scored at least 4. As a result, 98 papers were excluded and 34 studies were selected.  The result of the QA of the studies is presented in Table \ref{table:Quality} in Appendix A.

%% quality assessment criteria  QA5 %% citation taking into consideration the publishing year 

\end{description}

\begin{table}[!t]
\caption{Primary selection inclusion and exclusion criteria}
\centering
\resizebox{\textwidth}{!}{
\begin{tabular}{p{.25cm}p{5.8cm} p{.1cm} p{.25cm} p{5.8cm}}
\hline 

& \Centering\bfseries \textbf{Inclusion criteria}               && & \Centering\bfseries \textbf{Exclusion criteria} \\  \hline

a. & All papers published in the English language               && a. & Papers that are not published in the English language \\
b. & Papers related to at least one of the  research questions  && b. & Papers that do not have any link with the research questions  \\
c. & Relevant papers that are published from 1996 to 2016       && c. & If a paper has several versions only the most complete one is included \\

\hline
\end{tabular}}
\label{table:inexcriteria}
\end{table}

\begin{table}[!b]
\caption{Quality assessment questions}
\centering
\resizebox{0.99\textwidth}{!}{
\begin{tabular}{ll}
\hline 
\multicolumn{2}{l}{\textbf{Quality assessment questions}} \\  \hline

 %& Questions \\  \hline

Q1 & Are the objectives of the proposed work clearly justified? \\ 
Q2 & Are the proposed concepts/relations clearly defined? \\
Q3 & Does the work propose sufficient concepts/relations to deal with privacy aspects? \\
Q4 & Have the concepts/relations been applied to project/case study, or have they\\
   &  been justified by appropriate examples? \\
Q5 &  Does the work add value to the state-of-the-art\textsuperscript{1}? \\  %% citation taking into consideration the publishing year 

\hline
\multicolumn{2}{l}{\textsuperscript{1}Evaluated based on the number of citations taking into consideration the year of publication} \\  

\hline
\end{tabular}}
\label{table:qualityassessment}
\end{table}

\subsection{Reporting the results}

The final phase of the systematic review involves summarizing the results, and it consists of two main activities: 1- data synthesis; and 2- results and discussion.

\subsubsection{Data synthesis}

In what follows, we describe how data syntheses were executed: Data related to \textit{RQ1}  can be extracted directly from the list of  selected papers (shown in Table \ref{table:selected}). To answer \textit{RQ2}, the contents of the 34 selected studies were further analyzed to identify  privacy related concepts along with their interrelations, and list them in a comprehensive table (Table \ref{table:rq}). Moreover,  we identify  the main concepts/relations for capturing privacy requirements based on Table \ref{table:rq} \& Table \ref{table:iteration} that shows the frequency of concepts/relations appearance in the selected studies. To answer \textit{RQ3} data can be derived from Table \ref{table:limitation}, which summaries the percentage of the main concepts/relations categories that each selected study cover. \textit{RQ4} can be answered by categorizing the studies into four group based on the concepts categories they do not cover.

\section{Review results and discussion}

This section presents and discusses the findings of this review. First, we start by presenting an overview of the selected studies, and then, we present the findings of this review concerning the research questions. 

\begin{figure}[!h]
\centering
    \includegraphics[width=0.7\textwidth]{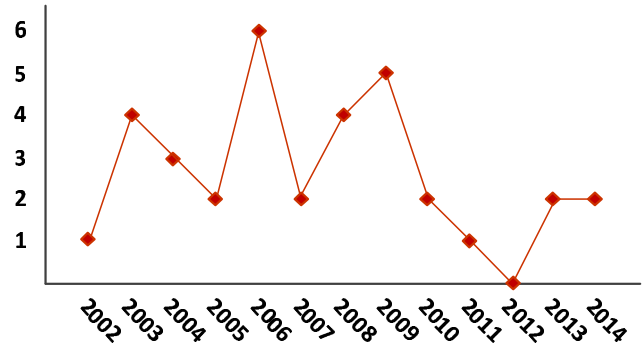}
    \caption{Number of papers by year of publication}
		\label{fig:pubyear}
\end{figure}

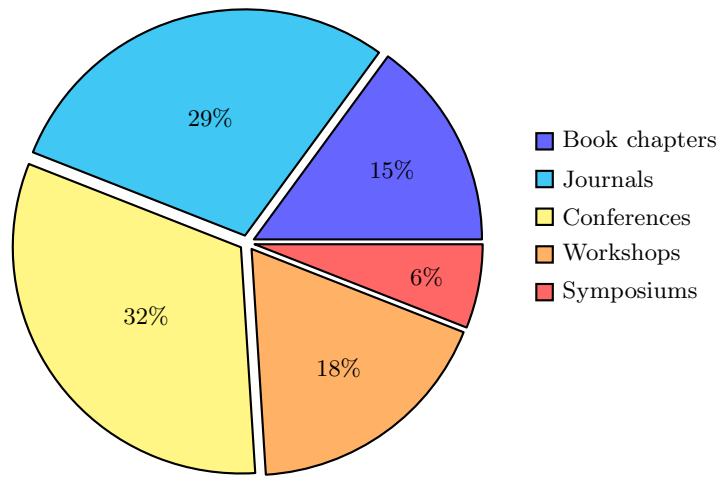
\begin{figure}[!b]
\centering
\begin{tikzpicture}
\pie[text = legend, explode ={ 0.1 , 0.1 , 0.1, 0.1, 0.1} ]{15/ Book chapters , 29/ Journals  , 32/ Conferences, 18/ Workshops , 6/Symposiums  }
\end{tikzpicture}
\caption{Percentages of selected studies} \label{fig:pie}
\end{figure}

\textbf{Overview of selected studies\footnote{An overview of all considered studies is shown in Table \ref{table:papers} in Appendix B}.} 34 studies were selected, where 5 studies were from book chapters, 10 papers were published in journals, 11 papers appeared in conference proceedings, 6 papers came from workshops, and 2 papers were extracted from symposiums.  The number of papers by year of publication is presented in Figure \ref{fig:pubyear}; while the percentages of the selected studies based on their publishing type are represented in Figure \ref{fig:pie}.

\begin{comment}
\begin{figure}[!h]
  \centering
  \begin{minipage}[b]{0.49\textwidth}
    \includegraphics[width=\textwidth]{pie.eps}
    \caption{Percentages of selected studies}
		\label{fig:pie}
  \end{minipage}
  \hfill
  \begin{minipage}[b]{0.49\textwidth}
    \includegraphics[width=\textwidth]{pubyear.eps}
    \caption{Number of papers by year of publication}
		\label{fig:pubyear}
  \end{minipage}
\end{figure}

\end{comment}

\textbf{RQ1:}  \textit{What are the privacy concepts/relations that have been used to capture privacy concerns?} The review has identified 34 studies that provide  concepts and relations that can be used for capturing privacy requirements. The list of the selected studies that answers our  first research question (\textit{RQ1}) is presented in Table \ref{table:selected}, where each paper is described by its identifier, title, author(s), publication year and number of citation. In what follows, we present  a summary of the contributions of each selected study.

\begin{comment}

\begin{figure}[!t]
\centering
\includegraphics[width= 0.99 \linewidth]{pie.eps}
\caption{Numbers and percentages of selected studies}
%\vskip -6pt
\label{fig:pie}
\end{figure}

\begin{figure}[!b]
\centering
\includegraphics[width= 0.99 \linewidth]{pubyear.eps}
\caption{Number of papers by year of publication.}
%\vskip -6pt
\label{fig:pubyear}
\end{figure}

\end{comment}

\begin{table*}[!t]
\caption{The list of the selected studies}
\centering
\resizebox{0.99\textwidth}{!}{
\begin{tabular}{  p{2.4cm} | p{9.9cm} | p{2.1cm} | p{0.6cm} | p{0.6cm}  }
%\hline 

\Centering\bfseries \textbf{Study ID} &  \Centering\bfseries \textbf{Title of the study} & \Centering\bfseries \textbf{Author(s)}   &  \Centering\bfseries  \textbf{Year} & \Centering\bfseries \textbf{\#Cite} \\  \hline

 \Centering ACM\_03 \cite{van2004elaborating}  &  Elaborating Security Requirements by Construction of Intentional Anti-Models & \Centering V. Lamsweerde &  \Centering 2004 & \Centering 337  \\ \hline

 \Centering ACM\_14 \cite{labda2014modeling} &   Modeling of privacy-aware business processes in BPMN to protect personal data   & \Centering Labda et al. &  \Centering 2014 & \Centering 0  \\  \hline

\Centering ACM\_16 \cite{braghin2008introducing} &   Introducing privacy in a hospital information system & \Centering	Braghin et al.	 &  \Centering 2008 & \Centering 9    \\  \hline

\Centering ACM\_35 \cite{singhal2010ontologies}  &   Ontologies for Modeling Enterprise Level Security Metrics  & \Centering A. Singhal, D. Wijesekera &  \Centering 2010 & \Centering 7   \\  \hline

\Centering ACM\_40 \cite{wang2009ovm}&   OVM: an ontology for vulnerability management    & \Centering  J. Wang and G. Minzhe  &  \Centering 2009 & \Centering 40 \\  \hline

\Centering IEEE\_12 \cite{souag2013using}  &   Using Security and Domain ontologies for Security Requirements Analysis & \Centering  Souag et al. &  \Centering 2013 & \Centering 4  \\  \hline

\Centering IEEE\_15 \cite{tsoumas2006towards}  &   Towards an Ontology-based Security Management  & \Centering B. Tsoumas, D. Gritzalis   &  \Centering 2006 & \Centering 88   \\  \hline  

\Centering IEEE\_50 \cite{Giorgini2005}  &   Modeling security requirements through ownership, permission and delegation & \Centering Giorgini et al. &  \Centering 2005 & \Centering 198  \\  \hline

\Centering IEEE\_57 \cite{kang2013security} &   A Security Ontology with MDA for Software Development & \Centering W. Kang, Y. Liang  &  \Centering 2013 & \Centering 1  \\  \hline  

\Centering CIT\_07 \cite{velasco2009modelling} &   Modeling Reusable Security Requirements Based on an Ontology Framework  & \Centering Lasheras et al.   & \Centering 2009 & \Centering 30  \\  \hline

\Centering CIT\_33 \cite{liu2003security}  &   Security and Privacy Requirements Analysis within a Social Setting & \Centering Liu et al.  & \Centering 2006 & \Centering 75 \\  \hline

\Centering Spgr\_07 \cite{massacci2011extended} &   An Extended Ontology for Security Requirements & \Centering  Massacci et al.   & \Centering 2011 & \Centering 16  \\  \hline

 \Centering Spgr\_13 \cite{elahi2009modeling} &   A Modeling Ontology for Integrating Vulnerabilities into Security Requirements Conceptual Foundation & \Centering Elahi et al. & \Centering 2009 & \Centering 21 \\  \hline

\Centering SCH\_18 \cite{sindre2005eliciting}&   Eliciting security requirements with misuse cases & \Centering G. Sindre and A. Opdahl & \Centering 2005 & \Centering 830  \\  \hline   

 \Centering SCH\_24 \cite{kalloniatis2008addressing} &   Addressing privacy requirements in system design: the PriS method & \Centering Kalloniatis et al. & \Centering 2008 & \Centering 76 \\  \hline

  \Centering SCH\_28 \cite{mouratidis2007secure} &   Secure Tropos: a security-oriented extension of the Tropos methodology & \Centering H. Mouratidis, P. Giorgini  & \Centering 2007 & \Centering 193  \\  \hline

 \Centering SCH\_41 \cite{solove2006taxonomy} &   A taxonomy of privacy & \Centering  D. Solove  & \Centering 2006 & \Centering 967 \\  \hline

\Centering Spgr\_18\_03 \cite{fenz2009formalizing} &   Formalizing information security knowledge & \Centering S. Fenz,  A. Ekelhart  & \Centering 2009 & \Centering 144  \\  \hline   

\Centering Spgr\_13\_01 \cite{asnar2008risk}  &   Risk as dependability metrics for the evaluation of business solutions: a model-driven approach & \Centering  Asnar et al.  & \Centering 2008 & \Centering 30  \\  \hline

\Centering Spgr\_13\_02 \cite{den2003coras} &   The CORAS methodology. model-based risk assessment using UML and UP & \Centering  Braber et al. & \Centering 2003 & \Centering 66  \\  \hline

\Centering Spgr\_13\_03 \cite{elahi2010vulnerability}  &   A vulnerability-centric requirements engineering framework. analyzing security attacks, countermeasures, and requirements based on vulnerabilities & \Centering  Elahi et al.  & \Centering 2010 & \Centering 73 \\  \hline

\Centering Spgr\_13\_04 \cite{jurjens2002umlsec}  &   UMLsec: Extending UML for secure systems development & \Centering J. J\"urjens  & \Centering 2002 & \Centering 583 \\  \hline

\Centering Spgr\_13\_05 \cite{matulevivcius2008adapting} &   Adapting Secure Tropos for security risk management in the early phases of information systems development & \Centering  Matulevi{\v{c}}ius et al. & \Centering 2008 & \Centering 60 \\  \hline

\Centering Spgr\_13\_07 \cite{rostad2006extended}  &   An extended misuse case notation: Including vulnerabilities and the insider threat & \Centering  L. R{\o}stad  & \Centering 2006 & \Centering 46  \\  \hline

\Centering Spgr\_08\_01 \cite{mayer2009model}  &   Model-based management of information system security risk & \Centering  N. Mayer  & \Centering 2009 & \Centering 70  \\  \hline

 \Centering Spgr\_08\_03 \cite{dritsas2006knowledge} &   A knowledge-based approach to security requirements for e-health applications & \Centering  Dritsas et al.  & \Centering 2006 & \Centering 17 \\  \hline

\Centering Spgr\_07\_02 \cite{zannone2006requirements} &   A requirements engineering methodology for trust, security, and privacy & \Centering  N. Zannone  & \Centering 2007 & \Centering 17  \\  \hline

\Centering Spgr\_07\_03 \cite{lin2003introducing} &   Introducing abuse frames for analysing security requirements & \Centering  Lin et al. & \Centering 2003 & \Centering 73  \\  \hline

 \Centering Spgr\_03\_01 \cite{avizienis2004basic}  &   Basic Concepts and Taxonomy of Dependable and Secure  & \Centering  Avi{\v{z}}ienis et al.  & \Centering 2004 & \Centering 3703 \\  \hline

\Centering Spgr\_02\_01 \cite{asnar2007trust}  &   From trust to dependability through risk analysis  & \Centering Asnar et al.  & \Centering 2007 & \Centering 57  \\  \hline

\Centering Spgr\_02\_02 \cite{asnar2006risk} &   Risk modelling and reasoning in goal models  & \Centering Asnar et al.  & \Centering 2006 & \Centering 17  \\  \hline

\Centering SCH\_24\_02 \cite{hong2004privacy} &   Privacy risk models for designing privacy-sensitive ubiquitous computing systems  & \Centering Hong et al.  & \Centering 2004 & \Centering 218 \\  \hline

\Centering SCH\_28\_01 \cite{paja2014sts} &   STS-Tool Security Requirements Engineering for Socio-Technical Systems  & \Centering  Paja et al.  & \Centering 2014 & \Centering 2  \\  \hline

 \Centering SCH\_43\_01 \cite{van2003handbook} &   Handbook of privacy and privacy-enhancing technologies & \Centering  Blarkom et al.  & \Centering 2003 & \Centering 69  \\

\hline
\end{tabular}}
\label{table:selected}
\end{table*}

\begin{description}
\item[\textbf{ACM\_03 \cite{van2004elaborating},}] ``\textit{Elaborating Security Requirements by Construction of Intentional Anti-Models}''.  Lamsweerde \cite{van2004elaborating} proposed a goal-oriented approach that extends the KAOS framework for modeling and analyzing security requirements. The framework focus on generating and resolving obstacles/anti-goals to goal satisfaction, i.e., it addresses malicious obstacles/anti-goals (threats) set up by attackers to threaten security goals, and the new security requirements are obtained as countermeasures to resolve these obstacles/anti-goals (threats).  The framework adopts several main concepts from KAOS (e.g., agents, goals, etc.) and proposes concepts for building intentional threat models (e.g., obstacles, anti-goal, anti-requirements, attacker, etc.). 

\item[\textbf{ACM\_14 \cite{labda2014modeling},}] ``\textit{Modeling of Privacy-aware Business Processes in BPMN to Protect Personal Data}''. Labda et al. \cite{labda2014modeling} propose a privacy-aware Business Processes (BP) framework for modeling, reasoning and enforcing privacy constraints. They have identified several privacy-related concepts, including: \textit{Data}, \textit{User}, \textit{Action}, \textit{Purpose}, and \textit{Permissions}.  In addition, they identify five concepts that can be used for analysis privacy in BP: (1) \textit{Access control}, (2) \textit{Separation of Tasks (SoT)}, (3) \textit{Binding of Tasks (BoT)}, (4) \textit{User consent}, (5) \textit{Necessity to know (NtK)}.

\item[\textbf{ACM\_16 \cite{braghin2008introducing},}] ``\textit{Introducing Privacy in a Hospital Information System}''.  Braghin et al. \cite{braghin2008introducing} presented an approach that supports expressing and enforcing privacy-related policies. The approach extends the conceptual model of an open source hospital information system (Care2x) with concepts for role-based privacy management (e.g., subject, processor, and controller), and concepts for supporting the privacy enforcement mechanisms (actions), where such actions can be either inactive or declarative, where the former includes actions that require to access and process data, while the latter includes simple statements representing activities that do not require to interact with the system.

\item[\textbf{ACM\_35 \cite{singhal2010ontologies},}] ``\textit{Ontologies for Modeling Enterprise Level Security Metrics}''. Singhal and Wijesekera \cite{singhal2010ontologies} provide a security ontology that supports IT security risk analysis. The ontology identifies which threats endanger which assets and what countermeasures can reduce the probability of the occurrence of a related attack. The concepts of the ontology, includes: \textit{threat}, a potential violation of security, an \textit{attack} exploits vulnerabilities to realize a threat, where \textit{vulnerabilities} are characteristics of target assets that make them prone to attack, and a \textit{risk} is an expectation of loss expressed as a probability that a particular threat will exploit a certain vulnerability, which will result in a harmful result. Finally, \textit{security mechanisms} are designed to prevent threats from happening or mitigating their impact when they occur.  

\item[\textbf{ACM\_40 \cite{wang2009ovm},}] ``\textit{OVM: An Ontology for Vulnerability Management}''. Wang and Guo \cite{wang2009ovm} propose an ontology for vulnerability management (OVM) that capture the fundamental concepts in information security and their relationship, retrieve vulnerable assets (data) and reason about the cause and impact of such vulnerabilities. The ontology has been built based on the Common Vulnerabilities and Exposures (CVE), Common Weakness Enumeration (CWE), Common Platform Enumeration (CPE), and Common Attack Pattern Enumeration and Classification (CAPEC). The top level concepts of the ontology includes, a \textit{Vulnerability} existing in an \textit{IT\_Product} that can be exploited by an \textit{Attacker} through an \textit{Attack} that compromises the \textit{IT\_Product} and cause \textit{Consequence}. Moreover, \textit{Countermeasures} can be used to protect the \textit{IT\_Product} through mitigating the \textit{Vulnerability}.

\item[\textbf{CIT\_07 \cite{velasco2009modelling},}] ``\textit{Modeling Reusable Security Requirements Based on an Ontology Framework}''.   Velasco et al. \cite{velasco2009modelling} propose an ontology-based framework for representing and reusing security requirements based on risk analysis. The ontology is based on two ontologies: 1- the risk analysis ontology that is developed based on MAGERIT \cite{magerit2006methodology}, and identifies five types of risk elements: \textit{asset},   \textit{threat}, \textit{safeguard}, \textit{valuation dimension}, \textit{valuation criteria}, and 2- the requirements ontology that models reusable requirements along with their relationships.

\item[\textbf{CIT\_33 \cite{liu2003security},}] ``\textit{Security and Privacy Requirements Analysis within a Social Setting}''.  Liu et al. \cite{liu2003security} propose a framework for dealing with security and privacy requirements within an agent-oriented modeling framework. They extend \textit{i}* modeling language to deal with security and privacy requirements, where \textit{i}* language allows for analyzing security/privacy issues within their social context, which enables for a systematic way of deriving vulnerabilities and threats. Moreover, \textit{i}* models make it possible to conduct different countermeasure analyses for addressing vulnerabilities and suggesting countermeasures for them.  

\item[\textbf{IEEE\_12 \cite{souag2013using},}] ``\textit{Using Security and Domain ontologies for Security Requirements Analysis}''. Souag et al. \cite{souag2013using} introduce an ontology-based method for discovering Security Requirements (SR). The process that underlies this method has three main steps, and it starts with the elicitation step that constructs an initial \textit{i}* requirements model from the stakeholders' needs/concerns about security.  The second step is the SR analysis that depends on production rules to exploit the security-specific ontology to discover threats, vulnerabilities, countermeasures, and resources. These concepts are used to enrich the requirements model by adding new elements (malicious tasks, vulnerability points, etc.). Finally, the domain specific SR analysis step, in which another set of rules explores the domain ontology to improve the requirements model with resources, actors and other concepts that are more specific to the domain at hand.

\item[\textbf{IEEE\_15 \cite{tsoumas2006towards},}] ``\textit{Towards an Ontology-based Security Management}''.  Tsoumas and Gritzalis \cite{tsoumas2006towards} introduce a security management framework that proposes a Security Ontology (SO), which contains the following concepts, a \textit{stakeholder} possesses an \textit{asset}, which in turn can be compromised by a \textit{vulnerability}. While a \textit{threat} initiated by a \textit{threat agent} targets an \textit{asset} and exploits a \textit{vulnerability} of the asset in order to achieve its goal. Exploitation of a \textit{vulnerability} leads to the realization of an unwanted \textit{incident}, which has a certain \textit{impact}. Furthermore, \textit{countermeasures} reduce the impact of the \textit{threat} with the use of \textit{controls}. Finally, \textit{security policy} formulates the \textit{controls} into a manageable security framework possessed by \textit{stakeholders}. 

\item[\textbf{IEEE\_50 \cite{Giorgini2005},}] ``\textit{Modeling Security Requirements through Ownership, Permission and Delegation}''.  Giorgini et al. \cite{Giorgini2005} introduce Secure Tropos, a formal framework for modeling and analyzing security requirements in their social and organizational context. Secure Tropos proposes several concepts including, an \textit{actor} that covers two concepts (a \textit{role} and an \textit{agent}), a \textit{goal} that can be refined through and/or-decompositions of a root \textit{goal} into finer \textit{sub-goals}, a \textit{task}, and a \textit{resource}. Secure Tropos adopts the notion of \textit{delegation} to model the transfer of objectives (\textit{goals} and \textit{tasks}) from one actor to another, and it adopts \textit{resource provision} among actors. Moreover, it introduces the \textit{ownership} concept that capture the relation between \textit{actors} and \textit{resources} they own. Finally, it provides the \textit{trust} concept to capture the \textit{actors'} expectations in one another concerning their social dependencies, and it introduce the \textit{monitoring} concept to compensate the lack of trust/distrust among \textit{actors} concerning social dependencies.

\item[\textbf{IEEE\_57 \cite{kang2013security},}] ``\textit{A Security Ontology with MDA for Software Development}''.  Kang and Liang \cite{kang2013security} propose security ontology for software development based on Model Driven Architecture (MDA) paradigm. The ontology includes most popular security concerns mentioned in literature such as \textit{auditing}, \textit{threats}, \textit{accountability}, \textit{non-repudiation}, \textit{risk}, \textit{attacks}, \textit{availability}, \textit{frauds}, \textit{confidentiality}, \textit{asset}, \textit{integrity}, \textit{prevention}, and \textit{Reputation}.

\item[\textbf{SCH\_18 \cite{kang2013security},}] ``\textit{Eliciting Security Requirements with Misuse Cases}''. Sindre and Opdahl \cite{kang2013security} present a systematic approach to eliciting security requirements based on \textit{use cases}. They extend the traditional \textit{use case} approach to also consider \textit{misuse cases} that represent unwanted behavior in the system to be developed.  In particular, a \textit{use case} diagram contains both, \textit{use cases} and \textit{actors}, as well as \textit{misuse cases} and \textit{misusers}.  In addition, \textit{misuse cases} adopts the ordinary \textit{use case} relationships such as \textit{include}, \textit{extend}, and \textit{generalize}. A \textit{use case} is related to a \textit{misuse case} using a directed \textit{association}, which means that a \textit{misuse case} \textit{threatens} the \textit{use case}.  Moreover, a use case diagram can contain \textit{security use cases}, which are special \textit{use cases} that can \textit{mitigate} \textit{misuse cases}. In summary, an ordinary \textit{use cases} represent requirements, \textit{security cases} represent security requirements, and \textit{misuse cases} represent security \textit{threats}.

\item[\textbf{SCH\_24 \cite{kalloniatis2008addressing},}] ``\textit{Addressing Privacy Requirements in System Design. the PriS Method}''. Kalloniatis et al. \cite{kalloniatis2008addressing} introduce PriS, a security requirements engineering method that consider users' privacy requirements. PriS considers privacy requirements as business goals and provides a methodological approach for analysing their effect onto the organizational processes.  The conceptual model of PriS is based on the Enterprise Knowledge Development (EKD) framework \cite{loucopoulos1999enterprise}, and it includes a set of concepts for modeling privacy requirements, such as: \textit{stakeholders}, \textit{goals} that can be either \textit{strategic goals} or \textit{operational goals}, and \textit{goals} can be \textit{realized} by \textit{processes}. On the other hand, \textit{privacy requirements} are a special type of \textit{goals} (\textit{privacy goals}), which constraint the causal transformation of organizational goals into processes. \textit{Privacy goals} may be decomposed in simpler goals or may \textit{support}/ \textit{conflict} the achievement of other \textit{goals}. Moreover, eight types of \textit{privacy goals} have been identified corresponding to the eight privacy concerns namely, authentication, authorisation, identification, data protection, anonymity, pseydonymity, unlinkability, and unobservability. 

\item[\textbf{SCH\_28 \cite{mouratidis2007secure},}] ``\textit{Secure Tropos: a Security-oriented Extension of the Tropos Methodology}''. Mouratidis and Giorgini \cite{mouratidis2007secure} introduce extensions to the Tropos methodology \cite{bresciani2004tropos} to model security concerns throughout the whole development process. Secure Tropos adopts from Tropos methodology concepts for modeling \textit{actors}, \textit{goals}, \textit{resources}, along with their different relations and social dependencies. In addition, it introduces concepts for modeling security requirements, such as a \textit{security constraint} (e.g., privacy, integrity, and availability), which can be decomposed into one or more security sub-constraints.  \textit{Security constraint} modeling is divided into \textit{security constraint delegation}, \textit{security constraint assignment}, and \textit{security constraint analysis}. Secure Tropos also introduces \textit{secure entity}, \textit{security features}, \textit{security mechanisms}, a \textit{secure capability}, a \textit{secure dependency}, and the \textit{threat} concept.

\item[\textbf{SCH\_41 \cite{solove2006taxonomy},}] ``\textit{A Taxonomy of Privacy}''. Solove \cite{solove2006taxonomy} provides taxonomy for understanding a wide range of privacy related problems. The taxonomy specifies four main groups of possible harmful activities:  \textbf{(i) information collection}: creates disruption based on the process of data gathering Two sub-classifications of information collection have been identified, \textit{surveillance} and  \textit{interrogation}. \textbf{(ii) information processing}: refers to the use, storage, and manipulation of data that has been collected. Five different sub-classifications of information processing have been identified: \textit{aggregation}, \textit{identification}, \textit{insecurity}, \textit{secondary use}, and \textit{Exclusion}.  \textbf{(iii) information dissemination}: in which the data holders transfer the information to others. Seven different sub-classifications of information dissemination have been identified:  \textit{breach of confidentiality}, \textit{disclosure}, \textit{exposure}, \textit{increased accessibility}, \textit{blackmail}, \textit{appropriation}, and 7- \textit{distortion}.  \textbf{(iv) invasion}: involves impingements directly on the individual. Two different sub-classifications of information invasion have been identified: \textit{intrusion} and 2-\textit{decisional interference}.

\item[\textbf{Spgr\_07 \cite{massacci2011extended},}] ``\textit{An Extended Ontology for Security Requirements}. Massacci et al. \cite{massacci2011extended} propose ontology for security requirements engineering, the ontology adopts concepts from Secure Tropos methodology \cite{massacci2007computer}, Problem Frame \cite{haley2008security}, and several industrial case studies. The most general concept in the ontology is \textit{Thing}. An \textit{object} is a \textit{thing} that persists, and an \textit{event} is an instantaneous happening that changes some \textit{objects}. The \textit{object} concept can be specialized into \textit{proposition}, \textit{situation}, \textit{entity} and \textit{relationship}. A \textit{proposition} is an \textit{object} representing a true/false statement. A \textit{situation} is a partial world described by a \textit{proposition}. An \textit{entity} is an \textit{object} that has a distinct, separate existence from all other \textit{things}, though that existence need not be material. \textit{Entity} is specialized into \textit{Actor}, \textit{Action}, \textit{Process}, \textit{Resource}, and \textit{Asset}.  \textit{Relationship} can be specialized into \textit{do-dependency}, \textit{can-dependency}, \textit{trust-dependency}, \textit{and/or} refinement, \textit{contributes}, \textit{provides}, \textit{uses}. In addition, \textit{damages} is a \textit{relationship} between an \textit{attack} and an \textit{asset}, where the \textit{attack} causes \textit{harm} to the \textit{asset}. \textit{Exploits} is a \textit{relationship} between \textit{attack} and \textit{vulnerability}. \textit{Protects} relates a \textit{security goal} to an \textit{asset}. Finally, \textit{denies} relates an \textit{anti-goal} to a \textit{requirement}. Finally, a \textit{specification} is an \textit{entity} consisting of \textit{actions}, quality propositions, and domain assumptions.     \textit{Vulnerability} is a specialization of Situation and is adopted from the Security domain. While a \textit{threat}consists of a situation that includes an attacker and one or more vulnerabilities.

\item[\textbf{Spgr\_13 \cite{elahi2009modeling},}] ``\textit{A Modeling Ontology for Integrating Vulnerabilities into Security Requirements Conceptual Foundation}''. Elahi et al. \cite{elahi2009modeling} propose a vulnerability-centric modeling ontology, which integrates empirical knowledge of vulnerabilities into the system development process. They identify a set of core concepts for security requirements elicitation, and they identify another set of concepts for capturing vulnerabilities and their effects on the system. The ontology contains several concepts, including: a \textit{concrete element} that is a tangible entity (e.g., an \textit{activity}, \textit{task}, etc.), and it may \textit{bring} a \textit{vulnerability} into the system. \textit{Exploitation} of \textit{vulnerabilities} can have effects on other elements (\textit{affected elements}), where the \textit{effect} relation is characterized by the \textit{severity} attribute. An \textit{attack} involves the execution of \textit{malicious actions} that one or more \textit{actors} perform to satisfy some \textit{malicious goal}.  A \textit{concrete element} may have a \textit{security impact} on \textit{attacks}, which can be interpreted as a \textit{security countermeasure} that can be used to patch \textit{vulnerabilities}. 

\item[\textbf{Spgr\_02\_01 \cite{asnar2007trust},}]  ``\textit{From Trust to Dependability Through Risk Analysis}''. Asnar et al. \cite{asnar2007trust} present an extension of the Tropos Goal-Risk framework. In particular, they introduce an approach to assess risk on the basis of trust relations among actors. In particular, they introduce the notion of trust to extend the risk assessment process. Using this framework, an actor can assess the risk in delegating the fulfillment of his objectives and decide whether or not the risk is acceptable. They also introduce the notion of trust level proposing three trust levels: \textit{Trust}, \textit{Distrust}, and \textit{NTrust} (i.e., neither trust nor distrust), where a low level of trust increases the risk perceived by the depender about the achievement of his objectives.

\item[\textbf{Spgr\_02\_02 \cite{asnar2006risk},}] ``\textit{Risk Modeling and Reasoning in Goal Models}''. Asnar et al. \cite{asnar2006risk} propose a goal-oriented approach for modeling and reasoning about risks at requirements level, where risks are introduced and analyzed along the stakeholders’ goals and countermeasures. Their proposed framework is based on the Tropos methodology and extends it with new concepts and qualitative reasoning mechanisms to consider risks since the early phases of the requirements analysis.  In their framework, a \textit{risk} is an \textit{event} that has a negative impact on the satisfaction of a \textit{goal}, while a \textit{treatment} is a \textit{countermeasure} that can be adopted in order to mitigate the effects of the \textit{risk}. Moreover, they consider \textit{likelihood} as a property of the \textit{event}, and they capture the \textit{likelihood} by the level of evidence that supports and prevents the occurrence of the \textit{event} (SAT and DEN). On the other hand, \textit{impact} is used to capture the influence of an event to the \textit{goal} fulfillment, and they classify impact under: \textit{strong positive}, \textit{positive}, \textit{negative}, and \textit{strong negative}.

\item[\textbf{Spgr\_03\_01 \cite{avizienis2004basic},}] ``\textit{Basic Concepts and Taxonomy of Dependable and Secure Computing}''. Avizienis et al. \cite{avizienis2004basic} propose a new taxonomy for dependable and secure computing based on an extensive analysis of the related literature.  The authors provide precise definitions characterizing the various concepts that come into play when addressing the dependability and security of computing and communication systems.  The three top-level dimensions of this taxonomy are: \textit{attribute}, \textit{threat}, and \textit{means}. The concept of \textit{attribute} is analyzed in terms of: \textit{availability}; \textit{reliability}; \textit{safety}; \textit{confidentiality}; \textit{integrity}; and \textit{maintainability}. The concept of \textit{threat} is further refined in terms of \textit{fault}, \textit{error}, and \textit{failure}.  While the concept of  \textit{means} is used to attain the various attributes of dependability and security, where these means can be grouped into four main categories: \textit{fault prevention}; \textit{fault tolerance}; \textit{fault removal}; and \textit{fault forecasting}.

\item[\textbf{Spgr\_07\_02 \cite{zannone2006requirements},}]  ``\textit{A Requirements Engineering Methodology for Trust, Security, and Privacy}''. Zannone \cite{zannone2006requirements} introduces the Secure i* (SI*) methodology that adopts from Secure Tropos the concepts of \textit{actors}, \textit{goals}, \textit{resources}, along with their different relations and social dependencies, and it proposes new relation among roles, namely \textit{supervision}. In SI*, an \textit{actor} is defined along with a set of \textit{objectives}, \textit{capabilities}, and \textit{entitlements}, which can be modeled through relations between actors and services (goals, tasks, and resources), namely: (1) \textit{require} indicates that an actor intends to achieve a \textit{service}, (2) \textit{be entitled} indicates that an \textit{actor} is the legitimate \textit{owner} of a \textit{service}, and (3) \textit{provide} indicates that the \textit{actor} has the capability to achieve a \textit{service}.   The delegation concept is refined in SI* into: (1) \textit{Delegation of execution (De)}, and (2) \textit{Delegation of permission (De)}. In addition, the trust concept is refined to cope with the refinement of delegation they propose into: (1) \textit{Trust of execution  (Te)}, and (2) \textit{Trust of permission (Tp)}. 

\item[\textbf{Spgr\_07\_03 \cite{lin2003introducing},}]  ``\textit{Introducing Abuse Frames for Analysing Security Requirements}''. Lin et al. \cite{lin2003introducing} develop an approach using Problem Frames to analyze security problems in order to determine security vulnerabilities.  In particular, they introduce the notion of an anti-requirement as the requirement of a malicious user that can subvert an existing requirement, and they incorporate anti-requirements into abuse frames to represent the notion of a security threat that is imposed by malicious users in a particular problem context. 

\item[\textbf{Spgr\_08\_01 \cite{mayer2009model},}] ``\textit{Model-based Management of Information System Security Risk}''. Mayer \cite{mayer2009model} proposes ISSRM (Information System Security Risk Management), a security risk management model.  The ISSRM reference model addresses risk management at three different levels, combining together \textit{asset}, \textit{risk}, and \textit{risk treatment} views, and it proposes concepts that are ordered in three main groups:  \textbf{(i) Asset-related concepts} describe what assets are important to protect, and what criteria guarantee asset security;  \textbf{(ii) Risk-related concepts} present how the risk itself is defined. A \textit{risk} is the combination of a \textit{threat} with one or more \textit{vulnerabilities} leading to a negative impact harming the \textit{assets}; and \textbf{(iii) Risk treatment-related concepts} describe what decisions, requirements and controls should be defined and implemented in order to mitigate possible \textit{risks}. 

\item[\textbf{Spgr\_08\_03 \cite{dritsas2006knowledge},}]  ``\textit{A Knowledge-based Approach to Security Requirements for E-health Applications}''. Dritsas et al. \cite{dritsas2006knowledge} propose an ontology that includes the main security related concepts, and use the ontology for designing and developing a set of security patterns that address a subset of these requirements for applications that provide e-health services.  The concepts used in the proposed ontology includes: \textit{stakeholder}, \textit{objective}, \textit{threat}, \textit{countermeasure}, \textit{asset}, \textit{vulnerability}, \textit{deliberate attack}, \textit{security pattern} and \textit{security pattern context}.  A \textit{security pattern} provides a specific set of \textit{countermeasures}, and a \textit{security pattern context} is defined as a set of \textit{asset}, \textit{vulnerability}, and \textit{deliberate attack} triplets. Therefore, one can start from the generic \textit{security objectives}, find the \textit{security pattern contexts} that match them and choose specific \textit{security pattern}, which ensures that the high level \textit{security objectives} can be fulfilled by implementing the respective \textit{countermeasures}. 

\item[\textbf{Spgr\_13\_01 \cite{asnar2008risk},}]    ``\textit{Risk as Dependability Metrics for the Evaluation of Business Solutions: a Model-driven Approach}''. Asnar et al. \cite{asnar2008risk} adopt and extend the Tropos Goal Model \cite{asnar2006risk,asnar2007trust} by considering also the interdependency among the actors within an organization. Through this extension analysts can assess the risk perceived by each actor, taking into account the organizational environment where the actor acts. Based on such analysis, we have provided a method to assist analysts in determining the treatments to be introduced in order to make risks be acceptable by all actors. 

\item[\textbf{Spgr\_13\_02 \cite{den2003coras},}]     ``\textit{The CORAS Methodology: Model-based Risk Assessment Using UML and UP}''. Braber \cite{den2003coras} introduces the CORAS methodology in which the Unified Modeling Language (UML) and Unified Process (UP) are combined to support a model-based risk assessment of security-critical systems.  The CORAS ontology propose several concepts, such as \textit{context} that influences the \textit{target}, which contains \textit{assets} and has its \textit{security requirements}. \textit{Security requirements} lead to \textit{security policies}, which protect \textit{assets} by reducing its related \textit{vulnerabilities} that can be exploited by \textit{threats}, which might reduce the \textit{value} of the \textit{asset}.  A \textit{Risk} contains an \textit{unwanted incident} that has a certain \textit{consequence} and \textit{frequency} of occurrence. 

\item[\textbf{Spgr\_13\_03 \cite{elahi2010vulnerability},}]   ``\textit{A Vulnerability-centric Requirements Engineering Framework: Analyzing Security Attacks, Countermeasures, and Requirements based on Vulnerabilities}. Elahi et al. \cite{elahi2010vulnerability} adopt and extend their previous work \cite{elahi2009modeling} by proposing an agent- and goal-oriented framework for eliciting and analyzing security requirements.  They refined the goal model evaluation method that helps analysts verifying whether top goals are satisfied with the risk of vulnerabilities and attacks and assess the efficacy of security countermeasures against such risks. More specifically, the evaluation does not only specify if the goals are satisfied, but also makes it possible to understand why and how the goals are satisfied (or denied) by tracing back the evaluation to vulnerabilities, attacks, and countermeasures.

\item[\textbf{Spgr\_13\_04 \cite{jurjens2002umlsec},}]  ``\textit{UMLsec: Extending UML for Secure Systems Development}''. J{\"u}rjens \cite{jurjens2002umlsec} proposes UMLsec that is an extension to UML modeling language, which allows for integrating security requirements modeling and analysis within the system development process. UMLsec is able to model security related features such as secrecy, integrity, access control, etc. It represents security feature on UML diagrams by providing several extension mechanisms, namely: (1) stereotypes: a new types of modeling elements that extends the semantics of existing types in the UML meta-model; (2) tagged values: that is used to associate data with model elements and (3) constraints: that are used to define criteria to determine whether requirements are met or not by the system design. In UMLsec, integrity is modeled as a constraint, which can restrict unwanted modification (e.g., insert), but information quality can be affected in several other ways that cannot be captured by this approach.

\item[\textbf{Spgr\_13\_05 \cite{matulevivcius2008adapting},}]   ``\textit{Adapting Secure Tropos for Security Risk Management in the Early Phases of Information Systems Development}''. Matulevi{\v{c}}ius  et  al. \cite{matulevivcius2008adapting} have analyzed how Secure Tropos can be applied to analyze security risks at the early IS development phases. Their analysis suggested a number of improvements for Secure Tropos in order to deal better with security risk management activities. In particular, Secure Tropos could be improved with additional constructs adopted from existing security risk management models (e.g., ISSRM (Information System Security Risk Management)) such as risk, risk treatment, and control. More specifically, among the suggested risk-related concepts is a risk that presents how the risk itself is defined, what are the major principles that should be taken into account when defining the possible risks. The risk is described by the cause of the risk, and the impact of the risk captures the potential negative consequence of the risk, which can be represented by a negative contribution link between the attack and the related security constraint, i.e., the impact negates the security criteria.

\item[\textbf{Spgr\_13\_07 \cite{rostad2006extended},}]   ``\textit{An Extended Misuse Case Notation: Including Vulnerabilities}''. R{\o}stad  \cite{rostad2006extended} proposes an extended misuse case notation that includes the ability to represent vulnerabilities and the insider threat. In particular, beside the main concepts of misuse case notation (e.g., \textit{actors}, \textit{use cases}, \textit{misuse cases}, \textit{misusers}, etc.). R{\o}stad  introduce the \textit{insider} concept to capture inside attackers,  since the \textit{misuser} concept in misuse cases was mainly proposed to address outside attackers. More specifically, an \textit{insider} is a \textit{misusers} that is also member of an authorized group for the entity being attacked. In addition, she introduce the \textit{vulnerability} concept that is a weakness in the system, which can be \textit{exploited} by the \textit{insider}.

\item[\textbf{Spgr\_18\_03 \cite{fenz2009formalizing},}]  ``\textit{Formalizing Information Security Knowledge}''. Fenz and Ekelhar \cite{fenz2009formalizing} introduce security ontology for information security domain knowledge. In their ontology, a \textit{vulnerability} is the absence of a proper safeguard,which could be exploited by a \textit{threat}. A \textit{threat} might threaten an \textit{asset}, and it can be \textit{exploited} by predefined \textit{threat}, and \textit{mitigation} is achieved by the implementation of one or more \textit{control}. In addition, the \textit{severity} of each \textit{vulnerability} is rated by a three-point scale (high, medium, and low). A \textit{threat} has a \textit{source}, and a related \textit{security objectives}. An \textit{asset} is categorized either as a tangible or an intangible asset. While the \textit{data} concept comprises meta-data on the knowledge of an organization. The \textit{person} concept is used to model physical \textit{persons} in the ontology, and the \textit{organization} concept comprises organizations in the broadest sense and assigns \textit{roles} to them. A \textit{role} is a physical person or organization relevant to the organization. Finally, a \textit{location} is used to relate location and threat information in order to assign a priori threat probabilities.

\item[\textbf{SCH\_24\_02 \cite{hong2004privacy},}]  ``\textit{Privacy Risk Models for Designing Privacy-sensitive Ubiquitous Computing Systems}''. Hong et al. \cite{hong2004privacy} propose a privacy risk model that captures privacy concerns at high abstraction level, and then refining them into concrete issues for specific applications.  The privacy risk model consists of two parts: (1) a \textit{privacy risk analysis} that  poses a series of questions to help designers think about the social and organizational context in which an application will be used, the technology used to implement that application, and control and feedback mechanisms that end-users will use; and (2) \textit{privacy risk management} that takes the unordered list of privacy risks from the privacy risk analysis, prioritizes them, and helps design teams identify solutions for helping end-users manage those issues.

\item[\textbf{SCH\_28\_01 \cite{paja2014sts},}]   ``\textit{STS-Tool: Security Requirements Engineering for Socio-Technical}''. Paja et al. \cite{paja2014sts} present the STS-Tool, a modeling and analysis support tool for STS-ml (Socio-Technical Security modeling language), a security requirements modeling language for socio-technical systems. STS-ml consists of three complementary views: 1- The social view, 2-  The information view, 3- The authorization view. Through these views, STS-ml supports different types of security needs: (1) \textit{Interaction (security) needs} are security-related constraints on goal delegations and document provisions; (2) \textit{Authorisation needs} determine which information can be used, how, for which purpose, and by whom; (3) \textit{Organisational constraints} constrain the adoption of roles and the uptake of responsibilities. In addition, STS-ml supports the following interaction security needs: 1. Over goal delegations:  (a) \textit{No-redelegation}, (b) \textit{Non-repudiation}, (c) \textit{Redundancy}, (d) \textit{Trustworthiness}, and (e) \textit{Availability}. 2. Over-document provisions: (a) \textit{Integrity of transmission}, (b) \textit{Confidentiality of transmission}, (c) \textit{Availability}. 3. From organizational constraints: (a) \textit{Separation of duties (SoD)}, and (b) \textit{Combination of duties (CoD)}.

\item[\textbf{SCH\_43\_01 \cite{van2003handbook},}]   ``\textit{Handbook of Privacy and Privacy-enhancing Technologies}''. Van Blarkom et al. \cite{van2003handbook} investigate several active areas related to privacy, Privacy-Enhancing Technologies (PET), intelligent software agents, and the inter-relations among these areas. Furthermore, they discussed the concepts of privacy and data protection, the European Directives that rule the protection of personal data and the relevant definitions. In particular, they investigate when personal data items become non-identifiable, the sensitivity of data, automated decisions, privacy preferences, and policies. In addition, they discussed existing technological solutions that offer agent user privacy protection, known under the name Privacy-Enhancing Technologies (PETs), the set of technologies/ principles that underlying PETs, and the legal basis for PET. Moreover, they discussed the Common Criteria for Information Technology Security Evaluation (CC) supplies important information for building privacy secure agents.

\end{description}

\textbf{RQ2:} \textit{What are the main concepts/relations that have been used for capturing privacy requirements?}   Each of the 34 selected studies has been deeply investigated to identify any concept/relation that can be used for capturing privacy requirements. We have focused on identifying any concept/relation that can be used for capturing privacy requirements in their social and organizational context.  More specifically, we tried to identify any concept that is related to privacy, social and organizational threats that might threaten privacy needs, treatment/countermeasures that can be used to mitigate threats concerning privacy needs. The result is shown in Table \ref{table:rq}, which presents the concepts/relations that have been identified in each selected studies. In particular,  55 concepts and relations\footnote{When there are more than one concept with very close meaning, we have chosen the most appropriate one to represent all} have been identified, which have been grouped into four main groups based on their types:

%\begin{landscape}
\setlength\tabcolsep{1pt}
\footnotesize
% Table generated by Excel2LaTeX from sheet 'Sheet1'
\begin{table}[htbp]
%\begin{table}[t!]
  \centering
\vspace*{-1cm}
  \caption{Summary of the privacy related concepts and relations identified in the studies}
    \hspace*{-58pt}
\begin{tabular}{>{\raggedright}c c c| p{0.3cm} |p{0.3cm} |p{0.3cm} |p{0.3cm} | p{0.3cm} |p{0.3cm} |p{0.3cm} |p{0.3cm} |p{0.3cm} |p{0.3cm} | p{0.3cm} |p{0.3cm} |p{0.3cm} |p{0.3cm} |p{0.3cm} |p{0.3cm} | p{0.3cm} |p{0.3cm} | p{0.3cm} |p{0.3cm} |p{0.3cm} | p{0.3cm} |p{0.3cm} | p{0.3cm} | p{0.3cm} |p{0.3cm} | p{0.3cm} |p{0.3cm} | p{0.3cm} | p{0.3cm} | p{0.3cm} |p{0.3cm} | p{0.3cm} |p{0.3cm} | p{0.3cm} |}
    %\toprule

\cline{4-37} &   &    & \rotatebox{90}{\cite{van2004elaborating}}    & \rotatebox{90}{\cite{labda2014modeling}}   & \rotatebox{90}{\cite{braghin2008introducing}}   & \rotatebox{90}{\cite{singhal2010ontologies}}   & \rotatebox{90}{\cite{wang2009ovm}} & \rotatebox{90}{\cite{velasco2009modelling}} & \rotatebox{90}{\cite{liu2003security}}  & \rotatebox{90}{\cite{souag2013using}} & \rotatebox{90}{\cite{tsoumas2006towards}} & \rotatebox{90}{\cite{Giorgini2005}} & \rotatebox{90}{\cite{kang2013security}}   & \rotatebox{90}{\cite{massacci2011extended}} & \rotatebox{90}{\cite{elahi2009modeling}} & \rotatebox{90}{\cite{sindre2005eliciting}} & \rotatebox{90}{\cite{kalloniatis2008addressing}}  & \rotatebox{90}{\cite{mouratidis2007secure}} & \rotatebox{90}{\cite{solove2006taxonomy}} & \rotatebox{90}{\cite{fenz2009formalizing}} & \rotatebox{90}{\cite{asnar2008risk}}   & \rotatebox{90}{\cite{den2003coras}}   & \rotatebox{90}{\cite{elahi2010vulnerability}}        & \rotatebox{90}{\cite{jurjens2002umlsec}}   & \rotatebox{90}{\cite{matulevivcius2008adapting}} & \rotatebox{90}{\cite{rostad2006extended}}      & \rotatebox{90}{\cite{mayer2009model}} & \rotatebox{90}{\cite{dritsas2006knowledge}}  & \rotatebox{90}{\cite{zannone2006requirements}} &         \rotatebox{90}{\cite{lin2003introducing}} & \rotatebox{90}{\cite{avizienis2004basic}} & \rotatebox{90}{\cite{asnar2007trust}} & \rotatebox{90}{\cite{asnar2006risk}} & \rotatebox{90}{\cite{hong2004privacy}}  & \rotatebox{90}{\cite{paja2014sts}} & \rotatebox{90}{\cite{van2003handbook}}   \\
\bottomrule
    \end{tabular}\hspace*{-58pt}%
  \label{table:rq}%
\end{table}%

\begin{description}
\item[\textbf{Organizational.}] 27 concepts and relations have been identified for capturing the agentive entities of the system in terms of their objectives, entitlements, dependencies and their expectations concerning such dependencies. The organizational concepts and relations are further grouped into four sub-categories:

\begin{description}
\item[\textbf{Agentive entities.}] 8 concepts and relations have been identified for capturing the active entities of the system (e.g., actor, user, etc. ).

\item[\textbf{Intentional entities.}] 5 concepts and relations have been identified for capturing objectives that active entities aim for achieve/want to perform (e.g., goal, task,  activity,  etc. ).

\item[\textbf{Informational entities.}] 8 concepts and relations have been identified for capturing informational assets  (e.g., data, asset, information, etc.).

\item[\textbf{Entities interactions.}] 6 concepts and relations have been identified for capturing the entities dependencies and expectations concerning such dependencies  (e.g., delegation, dependency, provision, trust, etc. ).
\end{description}
\end{description}

\begin{description}
\item[\textbf{Risk.}] 10 concepts and relations have been identified for capturing risk related aspects (e.g., risk, threat, vulnerabilities, attack, etc.).

\item[\textbf{Treatment.}] 8 concepts and relations have been identified for capturing treatment related aspects (e.g., treatment, countermeasure, mitigate etc.).

\item[\textbf{Privacy.}] 9 concepts and relations have been identified for capturing privacy related aspects (e.g., anonymity,  confidentiality, etc.).
\end{description}

Among the 55 identified concepts and relations, we have selected 38 main concepts and relations that can be used for capturing privacy requirements in their social and organizational context. In particular,  these  concepts and relations are 17 organizational, 9 risk, 5 treatment, and 7 privacy concepts, and they are shown in \textbf{Bold} typeset in Table \ref{table:rq}.  Each of the selected concepts and relations has been chosen based on the following two criteria: (1) its importance for capturing privacy requirements; and (2) the frequency of its appearance in the selected studies, which is shown in Table \ref{table:iteration}.

\textbf{RQ3:} \textit{Do existing privacy studies cover the main privacy concepts/relations?} We answer \textit{RQ3} by comparing the privacy related concepts/relations presented in each selected study with the main privacy concepts/relations identified while answering  \textit{RQ2}.  In Table \ref{table:rq}, we use (\checkmark) when the study presents a main privacy concept/relation, and (-) when the study presents a normal privacy concept/relation. In addition, we use (\text{\sffamily X}) to mark when a study misses a main  concept/relation. Table \ref{table:limitation} summarizes the percentage of the main privacy concepts/relations identified in each selected study with respect to the main four categories (organizational, risk, treatment, and privacy).  Considering Table \ref{table:rq} and Table  \ref{table:limitation}, it is easy to note that most studies miss main privacy related concepts/relations, i.e., none of them cover all the main  privacy related concepts/relations.  In \textbf{RQ4}, we discuss the limitation of each selected study.

\begin{table*}[!t]
\caption{The frequency of concepts/relations appearance in the selected studies}
\centering
\resizebox{0.99\textwidth}{!}{
\begin{tabular}{p{2.7cm} p{0.8cm} || p{2.7cm} p{0.8cm} || p{2.7cm} p{0.8cm} || p{2.7cm} p{0.8cm}  }
\hline 

 \textbf{Conc./rel.} & \textbf{\#} & \textbf{Conc./rel.} &  \textbf{\#} & \textbf{Conc./rel.} & \textbf{\#} & \textbf{Conc./rel.} &  \textbf{\#}  \\  \hline

\textbf{actor}       & \textbf{14} & \textbf{role}                & \textbf{10}   & \textbf{agent}           &  \textbf{12} & user                      & 5 \\ 

stakeholder          &  5          & person                       & 1             & \textbf{is\_a}           &  \textbf{7}  & \textbf{plays}            & \textbf{7} \\ 

\textbf{goal}        & \textbf{15} & objective                    & 3             & task                     & 12           & action                    & 7 \\ 

\textbf{refinement}  & \textbf{12} & asset                        & 14            & \textbf{information}     & \textbf{12}  & data                      & 6 \\ 

resource             &  8          & \textbf{personal info.}      & \textbf{5}    & sensitive info.          &  3           & \textbf{part\_of}         & \textbf{2} \\ 

\textbf{own}         &  \textbf{4} & \textbf{obj. deleg.}         & \textbf{7}    & \textbf{perm. deleg.}    &  \textbf{6}  & \textbf{info. provision}  & \textbf{6} \\ 

\textbf{monitor}     &  \textbf{4} & \textbf{obj. trust}          & \textbf{5}    & \textbf{perm. trust}     &  \textbf{3}  & risk                      & 11 \\ 

\textbf{threat}      & \textbf{16} & \textbf{intin. threat}       & \textbf{7}    & \textbf{casual threat}   &  \textbf{4}  & \textbf{vulnerability}    & \textbf{15} \\ 

attack               & 11          & \textbf{attacker}            &\textbf{16}    & \textbf{attack method}   &  \textbf{5}  & \textbf{impact}           & \textbf{10} \\ 

\textbf{threaten}    &  \textbf{8} & \textbf{exploits}            &\textbf{10}    & countermeasure           & 15           & \textbf{mitigate}         & \textbf{10} \\ 

control              &  6          & treatment                    & 1             & \textbf{s/p goal}        & \textbf{19}  & \textbf{s/p constraint}   & \textbf{4} \\ 

\textbf{s/p policy}  &  \textbf{4} & \textbf{s/p mechanism}       & \textbf{5}    & \textbf{sec/priv req.}   & \textbf{18}  & \textbf{confidentiality}  & \textbf{14} \\ 

integrity            & 13          & availability                 & 12            & \textbf{non-repudiation} & \textbf{3}   & \textbf{notice}           & \textbf{4} \\ 

\textbf{anonymity}   &  \textbf{4} & \textbf{transparency}        & \textbf{3}    & \textbf{accountability}  & \textbf{2}   &                           &   \\

\hline
\end{tabular}}
\label{table:iteration}
\end{table*}

\begin{table*}[!b]
\caption{The percentage of the main privacy concepts/relations captured by each selected study }
\centering
\resizebox{0.99\textwidth}{!}{
\begin{tabular}{p{2.4cm}  p{0.8cm}  p{0.8cm}  p{0.8cm} p{0.8cm} p{0.8cm} || p{2.4cm}  p{0.8cm}  p{0.8cm}  p{0.8cm} p{0.8cm} p{0.8cm}  }
\hline 

\textbf{Study}                             & \textbf{Org.} & \textbf{Risk}  & \textbf{Tre.} &  \textbf{Pri.} & \textbf{All} & \textbf{Study}                         & \textbf{Org.}  & \textbf{Risk}  & \textbf{Tre.}  &  \textbf{Pri.}   & \textbf{All}  \\  \hline

ACM\_03 \cite{van2004elaborating}               &      5/17     &   5/9          &      2/5      &    3/7    &   15/38  & ACM\_14 \cite{labda2014modeling}            &     2/17     &    0/9          &      0/5     &   2/7    &   4/38        \\ 

ACM\_16 \cite{braghin2008introducing}           &      3/17     &   0/9          &      2/5      &    0/7    &   5/38  & ACM\_35 \cite{singhal2010ontologies}         &      1/17     &    1/9          &      2/5      &    0/7   &   4/38        \\ 

ACM\_40 \cite{wang2009ovm}                      &      0/17     &    5/9         &      1/5      &    0/7    &  6/38  & CIT\_07 \cite{velasco2009modelling}           &     1/17      &   1/9           &      0/5      &    0/7    &  2/38         \\ 

CIT\_33 \cite{liu2003security}                  &    11/17      &   7/9          &      4/5      &    2/7    & 24/38  & IEEE\_12 \cite{souag2013using}                &     6/17      &   6/9           &      2/5      &    1/7    & 15/38         \\ 

IEEE\_15\cite{tsoumas2006towards}               &     0/17      &   5/9          &      3/5      &    2/7    & 10/38  & IEEE\_50 \cite{Giorgini2005}                  &    15/17      &   0/9           &      1/5      &    2/7    & 18/38         \\ 

IEEE\_57 \cite{kang2013security}                &     0/17      &   1/9          &      1/5      &    4/7    &  6/38  & Spgr\_7 \cite{massacci2011extended}           &     4/17      &   5/9           &      1/5      &    1/7    & 11/38         \\ 

Spgr\_13 \cite{elahi2009modeling}               &     2/17      &   6/9          &      1/5      &    1/7    & 10/38  & SCH\_18 \cite{sindre2005eliciting}            &     1/17      &   1/9           &      0/5      &    0/7    &   2/38         \\ 

SCH\_24 \cite{kalloniatis2008addressing}        &     0/17      &   1/9          &      1/5      &    0/7    &  2/38  & SCH\_28 \cite{mouratidis2007secure}           &     8/17      &   1/9           &      4/5      &    3/7    & 16/38         \\ 

SCH\_41 \cite{solove2006taxonomy}               &     5/17      &   0/9          &      0/5      &    3/7    &  8/38  & Spgr\_18\_03 \cite{fenz2009formalizing}       &     2/17      &   7/9           &      2/5      &    0/7    & 11/38         \\ 

Spgr\_13\_01 \cite{asnar2008risk}               &     3/17      &   0/9          &      1/5      &    0/7    &  4/38  & Spgr\_13\_02 \cite{den2003coras}              &     0/17      &   4/9           &      1/5      &    1/7    &  6/38         \\ 

Spgr\_13\_03 \cite{elahi2010vulnerability}      &    10/17      &   5/9          &      1/5      &    1/7    & 17/38  & Spgr\_13\_04 \cite{jurjens2002umlsec}         &     1/17      &   2/9           &      1/5      &    1/7    &   4/38          \\ 

Spgr\_13\_05 \cite{matulevivcius2008adapting}   &    10/17      &   4/9          &      2/5      &    1/7    & 17/38  & Spgr\_13\_07 \cite{rostad2006extended}         &    1/17      &   5/9           &      2/5      &    1/7    &  9/38          \\ 

Spgr\_08\_01 \cite{mayer2009model}              &     1/17      &   7/9          &      2/5      &    1/7    & 11/38  & Spgr\_08\_03 \cite{dritsas2006knowledge}       &    4/17      &   2/9           &      1/5      &    2/7    &  9/38          \\ 

Spgr\_07\_02 \cite{zannone2006requirements}     &     11/17     &   0/9          &      1/5      &    2/7    & 14/38  & Spgr\_07\_03 \cite{lin2003introducing}         &    0/17      &   4/9           &      0/5      &    1/7    &  5/38          \\ 

Spgr\_03\_01 \cite{avizienis2004basic}          &     0/17      &   1/9          &      0/5      &    1/7    &  2/38  & Spgr\_02\_01 \cite{asnar2007trust}             &    5/17      &   0/9           &      1/5      &    0/7    &  6/38          \\ 

Spgr\_02\_02 \cite{asnar2006risk}               &     2/17      &   0/9          &      1/5      &    0/7    &  3/38  & SCH\_24\_02 \cite{hong2004privacy}             &    2/17      &   3/9           &      0/5      &    3/7    &  8/38         \\ 

SCH\_28\_01 \cite{paja2014sts}                  &    15/17      &   1/9          &      1/5      &    5/7    & 22/38  & SCH\_43\_01 \cite{van2003handbook}             &    5/17      &   0/9           &      0/5      &    5/7    & 10/38         \\

\hline
\end{tabular}}
\label{table:limitation}
\end{table*}

% SCH\_41 missing ?  Solove \cite{solove2006taxonomy}

\textbf{RQ4:} \textit{RQ4 What are the limitations of existing privacy studies?}  We answer this question by categorizing the studies into four groups (\textbf{Group1-4}) \footnote{These groups are not mutually exclusive, i.e., a study may belong to all of them} based on the concepts categories (e.g., organizational, risk, treatment, and privacy) that the studies do not appropriately cover: 

\begin{description}

\item[\textbf{Group 1,}] contains  studies  that do not appropriately cover the organizational concepts. In this group, we have identified 25 studies out of the 34 selected ones, including:  ACM\_03 Lamsweerde \cite{van2004elaborating}, ACM\_14 Labda et al. \cite{labda2014modeling}, ACM\_16   Braghin et al. \cite{braghin2008introducing}, ACM\_35 Singhal and Wijesekera \cite{singhal2010ontologies}, ACM\_40  Wang and Guo \cite{wang2009ovm}, CIT\_07 Lasheras et al. \cite{velasco2009modelling}, IEEE\_12 Souag et al. \cite{souag2013using}, IEEE\_15 Tsoumas and Gritzalis \cite{tsoumas2006towards}, IEEE\_57 Kang and Liang \cite{kang2013security}, Spgr\_7 Massacci et al. \cite{massacci2011extended}, Spgr\_13 Elahi et al. \cite{elahi2009modeling}, SCH\_18 Sindre and Opdahl \cite{sindre2005eliciting}, SCH\_24 Kalloniatis et al. \cite{kalloniatis2008addressing}, Spgr\_18\_03 Fenz and Ekelhart \cite{fenz2009formalizing}, Spgr\_13\_01 Asnar et al. \cite{asnar2008risk}, Spgr\_13\_02  Braber et al. \cite{den2003coras}, Spgr\_13\_04  J\"urjens \cite{jurjens2002umlsec}, Spgr\_13\_07 R{\o}stad  \cite{rostad2006extended}, Spgr\_08\_01 Mayer \cite{mayer2009model},  Spgr\_08\_03 Dritsas et al. \cite{dritsas2006knowledge}, Spgr\_07\_03 Lin et al. \cite{lin2003introducing}, Spgr\_03\_01 Avi{\v{z}}ienis et al. \cite{avizienis2004basic}, Spgr\_02\_01 Asnar et al.  \cite{asnar2007trust},  Spgr\_02\_02 Asnar et al.  \cite{asnar2006risk}, SCH\_24\_02 Hong et al. \cite{hong2004privacy}, SCH\_43\_01 Blarkom et al. \cite{van2003handbook}.

\item[\textbf{Group 2,}] contains studies  that do not appropriately cover risk concepts. In this group, we have identified 22 studies out of the 34 selected ones, including:  ACM\_14 Labda et al. \cite{labda2014modeling}, ACM\_16 Braghin et al. \cite{braghin2008introducing}, ACM\_35 Singhal and Wijesekera \cite{singhal2010ontologies}, CIT\_07 Lasheras et al. \cite{velasco2009modelling},  IEEE\_50 Giorgini et al. \cite{Giorgini2005},  IEEE\_57 Kang and Liang \cite{kang2013security}, SCH\_18 Sindre and Opdahl \cite{sindre2005eliciting}, SCH\_24 Kalloniatis et al. \cite{kalloniatis2008addressing}, SCH\_28 Mouratidis and Giorgini \cite{mouratidis2007secure}, SCH\_41 Solove \cite{solove2006taxonomy}, Spgr\_13\_01 Asnar et al. \cite{asnar2008risk}, Spgr\_13\_02 Braber et al. \cite{den2003coras}, Spgr\_13\_04 J\"urjens \cite{jurjens2002umlsec}, Spgr\_08\_03 Dritsas et al. \cite{dritsas2006knowledge}, Spgr\_07\_02 Zannone \cite{zannone2006requirements}, Spgr\_07\_03 Lin et al. \cite{lin2003introducing}, Spgr\_03\_01 Avi{\v{z}}ienis et al. \cite{avizienis2004basic}, Spgr\_02\_01 Asnar et al.  \cite{asnar2007trust},  Spgr\_02\_02 Asnar et al.  \cite{asnar2006risk}, SCH\_24\_02 Hong et al. \cite{hong2004privacy}, SCH\_28\_01  Paja et al. \cite{paja2014sts}, SCH\_43\_01 Blarkom et al. \cite{van2003handbook}.

\item[\textbf{Group 3,}] contains studies  that do not appropriately cover treatment concepts. In this group, we have identified 31 studies out of the 34 selected ones, including: ACM\_03 Lamsweerde \cite{van2004elaborating}, ACM\_14 Labda et al. \cite{labda2014modeling}, ACM\_16 Braghin et al. \cite{braghin2008introducing}, ACM\_35 Singhal and Wijesekera \cite{singhal2010ontologies}, ACM\_40 Wang and Guo \cite{wang2009ovm}, CIT\_07 Lasheras et al. \cite{velasco2009modelling}, IEEE\_12 Souag et al. \cite{souag2013using},  IEEE\_50 Giorgini et al. \cite{Giorgini2005}, IEEE\_57 Kang and Liang \cite{kang2013security}, Spgr\_7 Massacci et al. \cite{massacci2011extended}, Spgr\_13 Elahi et al. \cite{elahi2009modeling}, SCH\_18 Sindre and Opdahl \cite{sindre2005eliciting}, SCH\_24 Kalloniatis et al. \cite{kalloniatis2008addressing}, SCH\_41 Solove \cite{solove2006taxonomy}, Spgr\_18\_03 Fenz and Ekelhart \cite{fenz2009formalizing}, Spgr\_13\_01 Asnar et al. \cite{asnar2008risk}, Spgr\_13\_02 Braber et al. \cite{den2003coras}, Spgr\_13\_03 Elahi et al. \cite{elahi2010vulnerability}, Spgr\_13\_04 J\"urjens \cite{jurjens2002umlsec}, Spgr\_13\_05 Matulevi{\v{c}}ius et al. \cite{matulevivcius2008adapting}, Spgr\_13\_07 R{\o}stad  \cite{rostad2006extended}, Spgr\_08\_01 Mayer \cite{mayer2009model}, Spgr\_08\_03 Dritsas et al. \cite{dritsas2006knowledge}, Spgr\_07\_02 Zannone \cite{zannone2006requirements},  Spgr\_07\_03 Lin et al. \cite{lin2003introducing}, Spgr\_03\_01 Avi{\v{z}}ienis et al. \cite{avizienis2004basic}, Spgr\_02\_01 Asnar et al.  \cite{asnar2007trust},  Spgr\_02\_02 Asnar et al.  \cite{asnar2006risk}, SCH\_24\_02 Hong et al. \cite{hong2004privacy}, SCH\_28\_01 Paja et al. \cite{paja2014sts}, SCH\_43\_01 Blarkom et al. \cite{van2003handbook}.

\item[\textbf{Group 4,}] contains studies  that do not appropriately cover the privacy concepts. In this group, we have identified 31 studies out of the 34 selected ones, including: ACM\_03 Lamsweerde \cite{van2004elaborating}, ACM\_14 Labda et al. \cite{labda2014modeling}, ACM\_16 Braghin et al. \cite{braghin2008introducing}, ACM\_35 Singhal and Wijesekera \cite{singhal2010ontologies}, ACM\_40 Wang and Guo \cite{wang2009ovm}, CIT\_07 Lasheras et al. \cite{velasco2009modelling}, CIT\_33 Liu et al. \cite{liu2003security}, IEEE\_12 Souag et al. \cite{souag2013using}, IEEE\_15 Tsoumas and Gritzalis \cite{tsoumas2006towards},  IEEE\_50 Giorgini et al. \cite{Giorgini2005},  Spgr\_7 Massacci et al. \cite{massacci2011extended}, Spgr\_13 Elahi et al. \cite{elahi2009modeling},  SCH\_18 Sindre and Opdahl \cite{sindre2005eliciting}, SCH\_24 Kalloniatis et al. \cite{kalloniatis2008addressing}, SCH\_28 Mouratidis and Giorgini \cite{mouratidis2007secure}, SCH\_41 Solove \cite{solove2006taxonomy}, Spgr\_18\_03 Fenz and Ekelhart \cite{fenz2009formalizing}, Spgr\_13\_01 Asnar et al. \cite{asnar2008risk}, Spgr\_13\_02 Braber et al. \cite{den2003coras},  Spgr\_13\_03 Elahi et al. \cite{elahi2010vulnerability}, Spgr\_13\_04 J\"urjens \cite{jurjens2002umlsec}, Spgr\_13\_05 Matulevi{\v{c}}ius et al. \cite{matulevivcius2008adapting}, Spgr\_13\_07 R{\o}stad  \cite{rostad2006extended}, Spgr\_08\_01 Mayer \cite{mayer2009model}, Spgr\_08\_03 Dritsas et al. \cite{dritsas2006knowledge}, Spgr\_07\_02 Zannone \cite{zannone2006requirements}, Spgr\_07\_03 Lin et al. \cite{lin2003introducing}, Spgr\_03\_01 Avi{\v{z}}ienis et al. \cite{avizienis2004basic}, Spgr\_02\_01 Asnar et al.  \cite{asnar2007trust},  Spgr\_02\_02 Asnar et al.  \cite{asnar2006risk}, SCH\_24\_02 Hong et al. \cite{hong2004privacy}. 

\end{description}

Based on the previous categories, we have 15 studies that do not appropriately cover all the four concepts categories, and 13 studies that do not appropriately cover three categories. 5 studies do not appropriately cover two categories,  and one study does not appropriately cover only one categories. A detailed description of the concepts and relations that each of these studies does not cover can be obtained from Table \ref{table:rq}.

Note that most of these studies have not been developed to address privacy related issues. Therefore, it is not a negative thing when they do not cover privacy related concepts. \textbf{RQ4} has been considered in this study to assist  authors of selected studies, if they aim to extend their frameworks and approaches to cover privacy concerns.

%============================================================================================================================

\section{A novel privacy ontology}

Several resent studies stress the need for addressing privacy concerns during the system design (e.g., Privacy by Design (PbD) \cite{kalloniatis2008addressing,labda2014modeling}). Nevertheless, based on the results of this review, it is easy to note that no existing study covers all the main privacy concepts/relations that have been identified in the review, i.e.,  no existing ontology enables for capturing main privacy aspects and without such ontology it is almost impossible to address privacy concerns during the system design. Therefore, proposing such ontology would be a viable solution for this problem. To this end, we propose a novel privacy ontology based on the main privacy concepts/relations identified in Table \ref{table:rq}.  The meta-model of our ontology is depicted in Figure \ref{fig:onto}, and the concepts of the ontology are organized into four main dimensions:

\begin{description}
\item[\textbf{Organizational dimension:}] proposes concepts to capture the social and technical components of the system in terms of their capabilities, objectives, and dependencies.

\item[\textbf{Risk dimension:}] proposes concepts to capture risks that might endanger privacy needs at the social and organizational levels.

\item[\textbf{Treatment dimension:}] proposes concepts to capture countermeasure techniques to mitigate risks to privacy needs.

\item[\textbf{Privacy dimension:}] proposes concepts to capture the stakeholders' (actors) privacy requirements/needs concerning their personal information.
\end{description}

In what follows, we define each of these dimensions in terms of their concepts and relations

\textbf{(1) Organizational dimension.} Most current complex systems consist of several autonomous components that interact and depend on one another for achieving their objectives. Therefore, this dimension includes the organizational concepts of the system, which have been further organized into several categories, including: intentional entities, entities' objectives, informational assets, entities interactions, and entities expectations concerning such interactions (social trust).   In what follows, we define each of these dimensions along with their concepts and relations. 

\underline{\textbf{Agentive entities:}} represent  the active entities of the system, we have selected three concepts along with two relations: 

\begin{description}
\item[\textbf{Actor}] represents an autonomous entity that has intentionality and strategic goals within the system. Actor can be decomposed into sub-units:

\item[\textbf{Role}] is an abstract characterization of an actor in terms of a set of behaviors and functionalities within some specialized context. A role can be a specialization (\textbf{is\_a}) of one another.

\item[\textbf{Agent}] is an autonomous entity that has a specific manifestation in the system. An agent can \textbf{plays} a role or more within the system, i.e., an agent inherits the properties of the roles it plays.
\end{description}

\underline{\textbf{Intentional entities:}} the behavior of actors is, usually, determined by the objectives they aim to achieve. Therefore, we adopted the goal concept and and/or decomposition (refinement) relations to represent such objectives. 

\begin{description}

\item[\textbf{Goal}] is a state of affairs that an actor intends to achieve.  When a goal is too coarse to be achieved, it can be refined through \textit{and/or-decompositions} of a root goal into finer sub-goals.  

\item[\textbf{And-decomposition}] implies that the achievement of the root-goal requires the achievement of all its sub-goals. 

\item[\textbf{Or-decomposition}] is used to provide different alternatives to achieve the root goal, and it implies that the achievement of the root-goal requires the achievement of any of its sub-goals. 
\end{description}

\underline{\textbf{Informational entities:}}  information is  one of the most important concepts when we speak about privacy. Among the available concepts for capturing informational asset, e.g., data \cite{labda2014modeling}, a resource (physical or informational) \cite{Giorgini2005,zannone2006requirements,mouratidis2007secure,massacci2011extended}, asset \cite{kang2013security,elahi2010vulnerability}, etc., we have adopted the following concepts and relations:

\begin{description}
\item[\textbf{Information}] represents any informational entity without intentionality. Information can be atomic or composite (composed of several parts), and we rely on \textit{part\_of} relation to capture the relation between an information entity and its sub-parts. In the context of this work, we differentiate between two main types of information:

\begin{description}
\item[\textbf{Personal information}] any information that can be \textit{related} (directly or indirectly) to an identified or identifiable legal entity (e.g., names, addresses, medical records, etc.), who has the right to control how such information can be used by others \cite{braghin2008introducing,van2003handbook}. 

\item[\textbf{Public information}] any information that cannot be \textit{related} (directly or indirectly) to an identified or identifiable legal entity, or personal information that has been made public by its legal entity \cite{labda2014modeling}.
\end{description}
\end{description}

\underline{\textbf{Information type of use:}} actors may use information to achieve their goals. Our ontology adopts three relations between goals and information(e.g., \textit{produce}, \textit{read}, and \textit{modify}), where each of these relations can be defined as follows:

\begin{description}
\item[\textbf{Produce}] indicates that information can be created by achieving the goal that is responsible for its production;
\item[\textbf{Read}] indicates that the goal achievement depends on consuming such information;
\item[\textbf{Modify}] indicates that the goal achievement depends on modifying such  information.
\end{description}

\underline{\textbf{Information ownership \& permissions:}} as previously mentioned, we differentiate between personal and public information if it can be \textit{related} (directly or indirectly) to an identified or identifiable legal entity. In what follows, we define the \textit{own} concept that relates personal information to its legal entity, and we specify how information owner controls  the usage to its personal information.

\begin{description}
\item[\textbf{Own}] indicates that an actor is the legitimate owner of information, where information owner has full control over the use of information it owns.
\end{description}

\begin{comment}
\begin{figure}[!ht]
\vspace*{-2cm}
%\hspace*{-2cm}
\centering
\includegraphics[width=  \textheight,angle=90]{onto.eps}
\caption{The proposed privacy ontology}
%\vskip -6pt
\label{fig:onto}
\end{figure}
\end{comment}

\noindent%
\begin{minipage}{\linewidth}% to keep image and caption on one page
\vspace*{-1cm}
\makebox[\linewidth]{%        to center the image
  \includegraphics[keepaspectratio=true,scale=0.99,angle=90]{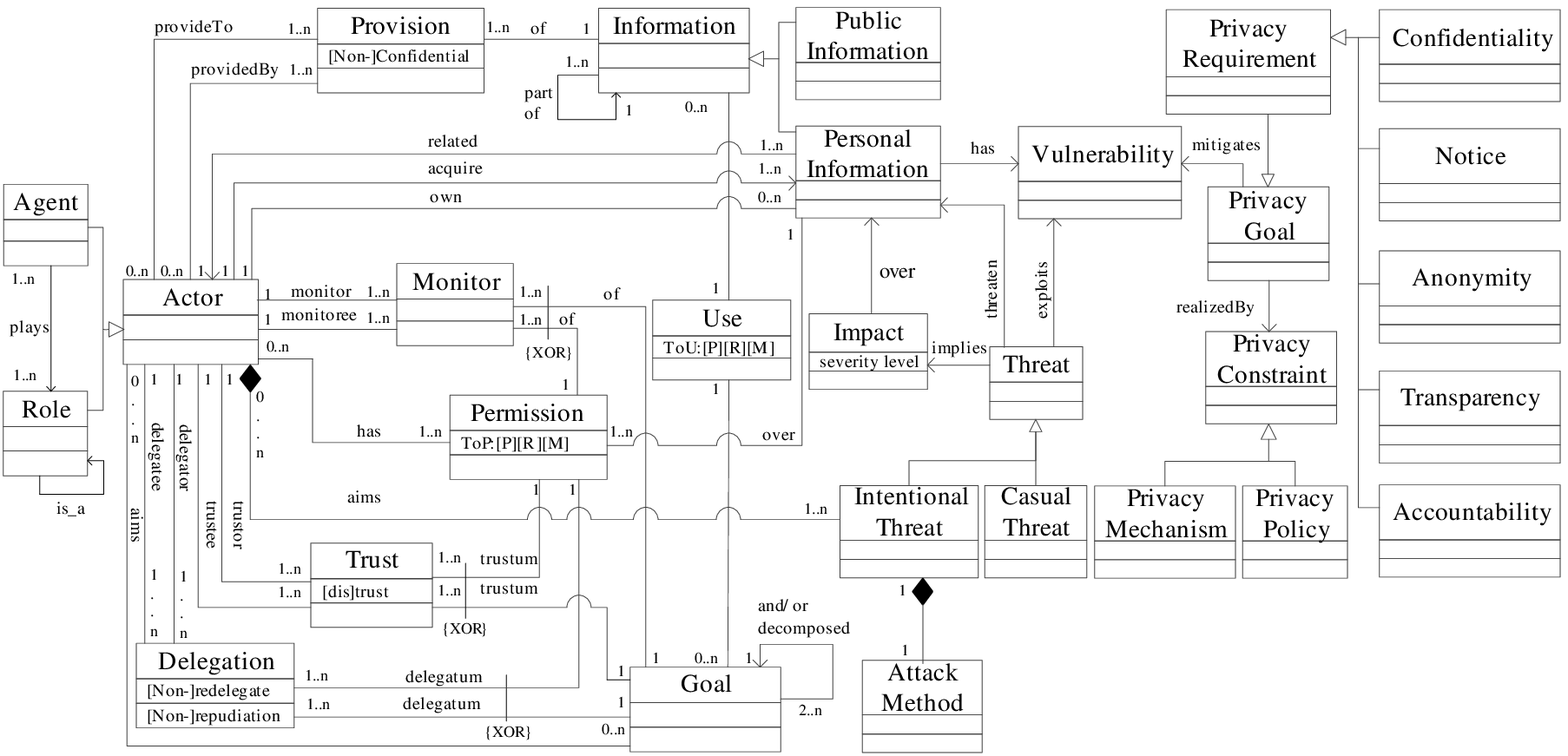}}
\captionof{figure}{The meta-model of the proposed privacy ontology}\label{fig:onto}%      only if needed  
\end{minipage}

\begin{description}
\item[\textbf{A permission}] is consent of a particular use of a particular object in a system \cite{sandhu1996role}, i.e., the holder of the permission is allowed to perform some action(s) in the system. Information owner has the authority to control the use of its own information, i.e., the owner can control the delegated permissions over information it owns. In our ontology, information permissions are classified under (P)roduce, (R)ead, (M)odify permissions, which covers the three relations between goals and information that our ontology propose. 
\end{description}

\underline{\textbf{Entities interactions:}} actors may not have the required capabilities to achieve their own objectives by themselves (e.g., achieve a goal, furnish information, etc.). Therefore, they depend on one another for such objectives. In what follows, we discuss the concepts that are used for capturing the different actors' social interactions and dependencies.

\begin{description}
\item[\textbf{Information provision}] indicates that an actor has the capability to deliver information   to another one, where the source of the provision relation is the provider and the destination is the requester. Information provision has one attribute that describes the provisioning type, which can be either \textit{confidential} or \textit{non-confidential}, where the first guarantee the confidentiality of the transmitted information while the last does not.
\end{description}

\begin{description}

\item[\textbf{Goal delegation}] indicates that one actor delegates the responsibility to achieve a goal to other actors, where the source of delegation called the delegator , the destination is called delegatee, and the subject of delegation is called delegatum.  

\item[\textbf{Permissions delegation}]  indicates that an actor delegates the permissions to produce, read and/or modify  over a specific information to another actor.

\end{description}

\underline{\textbf{Entities social trust:}} the need for trust arises when actors depend on one another for goals or permissions since such dependencies might entail risk  \cite{chopra2003trust,gharib2015analyzing}.  More specifically, a delegator has no warranty that the delegated goal will be achieved or the delegated permissions will not be misused by the delegatee. Therefore, our ontology adopts the notion of trust and distrust to capture the actors' expectations of one another concerning their delegations:

\begin{description}
\item[\textbf{Trust}] indicates the expectation of trustor that the trustee will behave as expected considering the trustum (e.g., trustee will achieve the delegated goal, or it will not misuse the delegated permission);

\item[\textbf{Distrust}] indicates the expectation of trustor that the trustee will not behave as expected considering the trustum (e.g., trustee will not achieve the delegated goal, or it will misuse the delegated permission). 
\end{description}

\underline{\textbf{Monitoring:}} we rely on monitoring to compensate the lack of trust or distrust in the trustee concerning the trustum \cite{gans2001modeling,zannone2006requirements}.
\begin{description}

\item[\textbf{Monitoring}] can be defined as the process of observing and analyzing the performance of an actor in order to detect any undesirable performance \cite{guessoum2004monitoring}, where the source of monitoring is called the monitor and the destination is called monitoree. 
\end{description}

\textbf{(2) Risk dimension.} Risk can be defined as an event that has a negative impact on the system, i.e., it is the possibility that a particular threat will harm one or more asset of a system by exploiting a vulnerability \cite{kang2013security,singhal2010ontologies,mayer2009model,elahi2009modeling}. In our ontology, risk is not a primitive concept and we do not include it into the ontology, since it can be captured by other concepts such as threat, vulnerabilities, attack, etc. In what follows, we define the risk dimension related concepts along with their interrelations: 

\begin{description}
\item[\textbf{A threat}] is a potential incident that \textit{threaten} an asset (personal information) by \textit{exploiting} a \textit{vulnerability} concerning such asset \cite{mayer2009model,singhal2010ontologies,kang2013security}. A \textit{threat} can be either natural (e.g. earthquake, etc.), accidental (e.g. hardware/software failure, etc.), or intentional (e.g. theft of personal information, etc.)\cite{fenz2009formalizing,velasco2009modelling,souag2015security}. Therefore, the ontology differentiates between two types of threat:

\item[\textbf{Casual threat}] (natural or accidental):   a threat that does not require a \textit{threat actor} nor an \textit{attack method}.

\item[\textbf{Intentional threat}] a threat that require a \textit{threat actor}  and a presumed \textit{attack method} \cite{lin2003introducing,massacci2011extended}. 

\item[\textbf{Threat actor}] is an actor that aims for achieving the \textit{intentional threat} \cite{rostad2006extended,mayer2009model,elahi2009modeling}.

\item[\textbf{Attack method}] is a standard means by which a \textit{threat actor} carries out an \textit{intentional threat} \cite{mayer2009model,elahi2010vulnerability,souag2015security}.

\item[\textbf{Impact}] is the consequence of the \textit{threat} \textit{over} the asset, and it can  be characterized by a \textit{severity} attribute that captures the level of the impact (e.g. high, medium or low) \cite{wang2009ovm,souag2015security}.

\item[\textbf{A vulnerability}] is a weakness in the system, asset (personal information), etc. that can be \textit{exploited} by a \textit{threat} \cite{rostad2006extended,mayer2009model,singhal2010ontologies}.\end{description}

\textbf{(3)  Treatment dimension.} This dimension introduces countermeasure concepts to mitigate risks, we adopted a high abstraction level countermeasure concepts to capture the required protection/treatment level (e.g., privacy goal), which can be refined into concrete protection/treatment constraints (e.g., mechanisms or policies) that can be implemented. The concepts of the treatment dimension are:
 
\begin{description}
\item[\textbf{A privacy goal}] is an aim to counter threats and prevents harm to personal information by satisfying privacy criteria concerning such information. 

\item[\textbf{A privacy constraint}] is a restriction that is used to realize/satisfy a privacy goal, constraints can be either a privacy policy or privacy mechanism.

\item[\textbf{A privacy policy}]  is a privacy statement that defines the permitted and/or forbidden actions to be carried out by actors of the system toward information.

\item[\textbf{A privacy mechanism}] is a concrete technique to be implemented for helping towards the satisfaction of privacy goal (attribute). 
\end{description}

\textbf{(4)  Privacy dimension.} Introduce concepts to capture the stakeholders' (actors) privacy requirements/needs concerning their personal information. The concepts of the privacy dimension are:

\begin{description}
\item[\textbf{Privacy requirement}]  is used to capture the actors' (personal information owner/subject) privacy needs at a high abstraction level, and it is specialized from the \textit{privacy goal} concept. Moreover, privacy requirement concept is further specialized into five more refined concepts. 

\item[\textbf{Confidentiality,}] means personal information should be kept secure from any potential leaks and improper access \cite{solove2006taxonomy,dritsas2006knowledge,labda2014modeling}. We rely on the following principles to analyze confidentiality:

\begin{description}
\item[\textbf{Non-disclosure,}] personal information can only be disclosed if the owner's consent is provided, i.e., the disclosure of the personal information should be under the control of its legitimate owner \cite{solove2006taxonomy,dritsas2006knowledge,braghin2008introducing,labda2014modeling}.  Note that \textit{non-disclosure} also cover information transmission that is why we differentiate between two types of information provision (confidential and non-confidential).

\item[\textbf{Need to know (NtK),}] an actor should only use information if it is strictly necessary for completing a certain task \cite{labda2014modeling,paja2014sts}.  

\item[\textbf{Purpose of use,}] personal information should only be used for specific, explicit, legitimate purposes and not further used  in a way that is incompatible with those purposes \cite{van2003handbook,solove2006taxonomy,dritsas2006knowledge}. \textit{Purpose of use} is able to address situations where an actor might be granted a permission to use some personal information for a legitimate purpose, yet after accessing it, he/she might use the information for some other purpose.
\end{description}

\item[\textbf{Notice,}] the data subject (information owner) should be notified when its information is being collected \cite{van2003handbook,solove2006taxonomy,dritsas2006knowledge}. Notice is considered mainly to address situations where personal information \textit{related} to a legitimate entity (data subject) is being collected without his/her knowledge. 

\item[\textbf{Anonymity,}] the identity of the information owner should not be disclosed unless it is required \cite{dritsas2006knowledge,solove2006taxonomy}, i.e., the primary/secondary identifiers of the data subject (e.g., name, social security number, address, etc. ) should be removed if they are not required and information still can be used for the same purpose after their removal. We rely on \textit{part\_of} relation to model the internal structure of personal information, i.e., we link the identifiers of the data subject with the rest of the information item by the \textit{part\_of} relation. If the identifiers are not required for the task, they can be easily removed, and information can be used without linking it back to its owner/data subject (unlinkability).

\item[\textbf{Transparency,}] information owner should be able to know who is using his/her information and for what purposes \cite{van2003handbook,dritsas2006knowledge,kang2013security}. We rely on the following principles to analyze transparency:

\begin{description}
\item[\textbf{Authentication,}] a mechanism that aims at verifying whether actors are who they claim they are \cite{paja2014sts}.
\item[\textbf{Authorization,}] a mechanism that aims at verifying whether actors can use information in accordance with their credentials \cite{dritsas2006knowledge}.
\end{description}

\item[\textbf{Accountability,}] information owner should have a mechanism available to them to hold information users accountable for their actions concerning information  \cite{dritsas2006knowledge,kang2013security}. We rely on the following principles to analyze accountability:

\begin{description}
\item[\textbf{Non-repudiation,}] the delegator cannot repudiate he/she delegated; and the delegatee cannot repudiate he/she accepted the delegation \cite{kang2013security,paja2014sts}.
\item[\textbf{Not-re-delegation,}] the delegatee is requested by the delegator not to re-delegate the delegatum, i.e., the re-delegation of a goal/permission is forbidden \cite{paja2014sts}.
\end{description}
\end{description}

\section {Threats to validity}

After presenting and discussing our systematic literature review, we discuss the threats to its validity in this section. Following Runeson et al. \cite{runeson2009guidelines}, we classify threats to validity under  four types: construct, internal, external and reliability:

\textbf{1- Construct threats:} is concerned with to what extent a test measures what it claims to be measuring \cite{runeson2009guidelines}. Construct validity is particularly important, since it might influence the internal validity as well \cite{mackenzie2003dangers}. We have identified the following threats:

\begin{description}

\item[Poor conceptualization of the construct:] occurs when the predicted outcome of the study is defined too narrowly \cite{mackenzie2003dangers}, i.e., using only one factor to analyze the subject of the study.  To avoid this threat, the research objective has been transformed into several research questions and for each of these questions, several factors  were specified to evaluate whether they have been properly answered. In addition, we followed the best practices in the area to define the criteria while searching for and selecting the related studies (e.g.,  inclusion and exclusion criteria, quality assessment criteria, etc.).

\item[Systematic error:] may occur while designing  and conducting the review. To avoid such threat, the review protocol has been carefully designed based on well-adopted methods, and it has been strictly followed during the different phases of the review.
 \end{description}

\textbf{2- Internal threats:} is concerned with factors that have not been considered in the study, and they could have influenced the investigated factors in the study \cite{trochim2006research,runeson2009guidelines}. One internal threat has been identified:

\begin{description}
\item[Publication bias:] publication bias is a common threat to the validity of systematic reviews, and it refers to a situation where positive research results are more likely to be reported than negatives ones \cite{keele2007guidelines}. Our review focused on finding privacy related concepts/relations by reviewing the related literature, and there are no positive nor negative research results in such case. Despite this, we have specified very clear inclusion and exclusion criteria, and quality assessment criteria while searching for and selecting the related studies.

\end{description}

\textbf{3- External threats:} is concerned with to what extent the results of the study can be generalized \cite{runeson2009guidelines}.  One internal threat has been identified:

\begin{description}
\item[Completeness:] it is almost impossible to capture all related studies, yet our review protocol and search strategy were very carefully designed to cover as much as possible of the related studies. In addition, we might exclude some relevant non-English published studies since we only considered English studies in this review. To mitigate this limitation we performed a manual scan of the references of all the primary selected studies in order to identify those studies that were missed during the first search stage. However, we cannot guarantee that we have identified all the main available studies, which can be used to answer our research questions.
\end{description}

\textbf{4- Reliability threats:} is concerned with to what extent the study is dependent on the researcher(s), i.e., if another researcher(s) conducted the same study, the result should be the same. The search terms, search sources, inclusion and exclusion criteria, quality assessment questions, etc. are all available, and any researcher can repeat the review and he should get similar results.  However, the researcher should take into consideration the time when the studies search process was performed, i.e., the researcher should limit the search time to March 2016.

\section{Related work}

There are few systematic reviews concerning privacy/securities ontologies. For instance, Souag et al. \cite{souag2012ontologies} performed a systematic review that proposes an analysis and a typology of existing security ontologies. While Blanco et al. \cite{blanco2008systematic} conducted a systematic review with a main aim for identifying, extracting and analyzing the main proposals for security ontologies. Fabian et al. \cite{fabian2010comparison} present a conceptual framework for security requirements engineering by mapping the diverse terminologies of different security requirements engineering methods to that framework. Moreover, a security ontology for capturing security requirements have been presented in \cite{souag2015security}. However, the focus of all the previously mentioned studies was security ontology.

On the other hand, Blanco et al. \cite{blanco2011basis} conduct a systematic review for extracting the key requirements that an integrated and unified security ontology should have. While Mellado et al. \cite{mellado2010systematic} carried out a systematic review of the existing literature concerning security requirements engineering in order to summarize the current contributions and to provide a road map for future  research in this area. Iankoulova and Daneva \cite{iankoulova2012cloud} performed a systematic review concerning the security requirements of cloud computing. In particular, they have classified the main identified security requirements under nine sub-areas: access control, attack/harm detection, non-repudiation, integrity, security auditing, physical protection, privacy, recovery, and prosecution. Li \cite{li2011empirical} conducted a systematic review concerning online information privacy concerns, consequences, and moderating effects. Based on the review outcome, he proposed a framework to illustrate the relationships between the previously mentioned factors and to highlight opportunities for further improvement.  Finally, Fern{\'a}ndez-Alem{\'a}n  et al. \cite{fernandez2013security} performed systematic literature review for  identifying and analyzing critical privacy and security aspects of the electronic health record systems.

\section{Conclusions and Future Work}

In this paper, we argued that many wrong design decisions might be made due to the insufficient knowledge about the privacy-related concepts, and we advocate that a well-defined privacy ontology that captures the privacy related concepts along with their interrelations can solve this problem. Therefore, we conduct a systematic review concerning the existing privacy/security literature with a main purpose of identifying the main concepts along with their interrelation for capturing privacy requirements.  The objectives of the research were considered to have been achieved since the research questions posed have been answered. Moreover, we used the identified concepts/relations for proposing a privacy ontology to be used by software engineers while dealing with privacy requirements.

For future work, we aim to develop core privacy ontology to be used by software/security engineers when dealing with privacy requirements. To achieve that, we are planning to contact the authors of the selected studies to get their feedback concerning the proposed privacy ontology. In addition, we will conduct a controlled experiment with software/security engineers to evaluate the usability of the ontology. Finally, we plan to evaluate the completeness and validity of the ontology by deploying it to capture the privacy requirements for two real case studies that belong to different domains (e.g., medical sector and public administration).

%\section*{Acknowledgment}

%This research was partially supported by the ERC advanced grant 267856, ``Lucretius: Foundations for Software Evolution'', \url{http://www.lucretius.eu/}.

%\bibliographystyle{splncs}

\newpage

\section*{Appendix A: Quality assessment application}

\begin{longtable}{ | p{0.6cm} | p{2.3cm} | p{0.5cm} | p{0.5cm} | p{0.5cm} | p{0.5cm} | p{0.5cm} | p{0.3cm} | p{0.6cm} | p{2.3cm} | p{0.5cm} | p{0.5cm} | p{0.5cm} | p{0.5cm} | p{0.5cm} | p{0.3cm} | }

\caption{Quality assessment (selection stage 2) application} 
\label{table:Quality}

\endfirsthead
\endhead

\hline

\Centering\bfseries N & \Centering\bfseries ID & \Centering\bfseries Q1  & \Centering\bfseries Q2  & \Centering\bfseries Q3  & \Centering\bfseries Q4  & \Centering\bfseries Q5  &  \Centering\bfseries S. & \Centering\bfseries N & \Centering\bfseries ID & \Centering\bfseries Q1  & \Centering\bfseries Q2  & \Centering\bfseries Q3  & \Centering\bfseries Q4  & \Centering\bfseries Q5  &  \Centering\bfseries S \\ \hline

\Centering\bfseries \textbf{1} & ACM\_02 \cite{chase2009improving} & \Centering\bfseries \textbf{-}  & \Centering\bfseries \textbf{-}  & \Centering\bfseries \textbf{-}  & \Centering\bfseries \textbf{1}  & \Centering\bfseries \textbf{1}  &  \Centering\bfseries \textbf{2}   &

\Centering\bfseries \textbf{2} & ACM\_03 \cite{van2004elaborating} & \Centering\bfseries \textbf{1}  & \Centering\bfseries \textbf{1}  & \Centering\bfseries \textbf{1}  & \Centering\bfseries \textbf{1}  & \Centering\bfseries \textbf{1}  &  \Centering\bfseries \textbf{5}  \\

\Centering\bfseries \textbf{3} & ACM\_04 \cite{rezgui2002preserving} & \Centering\bfseries \textbf{-}  & \Centering\bfseries \textbf{-}  & \Centering\bfseries \textbf{-}  & \Centering\bfseries \textbf{1}  & \Centering\bfseries \textbf{1}  &  \Centering\bfseries \textbf{2}  &

\Centering\bfseries \textbf{4} & ACM\_05 \cite{kost2012privacy} & \Centering\bfseries \textbf{1}  & \Centering\bfseries \textbf{1}  & \Centering\bfseries \textbf{-}  & \Centering\bfseries \textbf{-}  & \Centering\bfseries \textbf{-}  &  \Centering\bfseries \textbf{2}  \\
\hline

\end{longtable}

\newpage

\section*{Appendix B: Overview of all the considered studies}  %% RQ1

\begin{longtable}{ | p{0.6cm} | p{1.7cm} | p{3cm} | p{2.9cm} | p{1cm} | p{0.9cm} | p{1.7cm} | }

\caption{Overview of all the considered studies} 
\label{table:papers}

\endfirsthead
\endhead

\hline

%\rowcolor{black!30}

\Centering\bfseries \textbf{N} & \Centering\bfseries \textbf{ID} &  \Centering\bfseries \textbf{Title} & \Centering\bfseries \textbf{Author(s)}   &  \Centering\bfseries  \textbf{Pub Year} & \Centering\bfseries \textbf{\# Cited} & \Centering\bfseries \textbf{Decision} \\  \hline

\Centering 001 & \Centering ACM\_01 \cite{olivier2002database} &  \Centering Database Privacy, Balancing Conﬁdentiality, Integrity and Availability & \Centering Martin S Olivier &  \Centering 2002 & \Centering 30 &  \Centering Excluded  stage 1  \\  \hline

\Centering 002 & \Centering ACM\_02 \cite{chase2009improving} &  \Centering Improving privacy and security in multi-authority attribute-based encryption & \Centering Melissa Chase, Sherman S.M. Chow &  \Centering 2009 & \Centering 375 &  \Centering Excluded  stage 2  \\ \hline

\Centering 003 & \Centering ACM\_03 \cite{van2004elaborating}  &  \Centering Elaborating Security Requirements by Construction of Intentional Anti-Models & \Centering Axel van Lamsweerde &  \Centering 2004 & \Centering 337 &  \Centering Selected   \\ \hline

\Centering 004 & \Centering ACM\_04 \cite{rezgui2002preserving} &  \Centering Preserving Privacy in Web Services  & \Centering Abdelmounaam Rezgui, Mourad Ouzzani, Athman Bouguettaya, Medjahed Brahim  &  \Centering 2002 & \Centering 102 &  \Centering Excluded  stage 2  \\  \hline

\Centering 005 & \Centering ACM\_05 \cite{kost2012privacy}  &  \Centering Privacy analysis using ontologies  & \Centering Martin Kost, Johann Christoph Freytag &  \Centering 2012 & \Centering 9 &  \Centering Excluded  stage 2  \\  \hline

\Centering 006 & \Centering ACM\_06 \cite{gandhi2011discovering} &  \Centering Discovering Multidimensional Correlations among Regulatory Requirements to Understand Risk  & \Centering Robin A. Gandhi, Seok Won Lee   &  \Centering 2011 & \Centering 7 &  \Centering Excluded  stage 2  \\
\hline

\end{longtable}

\end{document}